\documentclass[12pt, review,authoryear]{elsarticle}
\usepackage{graphics}
\usepackage{graphicx}
\usepackage{epstopdf}
\usepackage{amsmath}
\usepackage{amssymb}
\usepackage{subfig}
\usepackage{float}
\usepackage{color}

\usepackage{natbib}

\graphicspath{{../ink/}, {../figs/}}

\makeatletter
\providecommand*{\diff}%
    {\@ifnextchar^{\DIfF}{\DIfF^{}}}
\def\DIfF^#1{%
    \mathop{\mathrm{\mathstrut d}}%
        \nolimits^{#1}\gobblespace}
\def\gobblespace{%
    \futurelet\diffarg\opspace}
\def\opspace{%
    \let\DiffSpace\!%
    \ifx\diffarg(%
        \let\DiffSpace\relax
    \else
        \ifx\diffarg[%
            \let\DiffSpace\relax
    \else
        \ifx\diffarg\{%
            \let\DiffSpace\relax
        \fi\fi\fi\DiffSpace}

\providecommand*{\deriv}[3][]{%
    \frac{\diff^{#1}#2}{\diff #3^{#1}}}

\newcommand{\tten}[1]{\times 10^{#1}}

\newcommand{\etal}{\emph{et al.~}}
\newcommand{\unit}[1]{$^{#1}$}
\newcommand{\sub}[1]{$_{\rm #1}$}

\newcommand{\bbeta}{B}%{\bar{\beta}}
\newcommand{\betaNO}{\beta_{NO}}
\newcommand{\bbetaNO}{\bbeta_{NO}}

\newcommand{\rSG}{\frac{sg}{1+\kappa_{SG}s}}
\newcommand{\rSC}{\frac{sm_c}{1+\kappa_{SC}s}}

\newcommand{\rNO}{\frac{nm_o}{1+\kappa_{NO}n}}

\newcommand{\ccarb}{\left(1+\frac{\nu_Cg}{\omega_C+g}\right)}
\newcommand{\ocarb}{\left(1+\frac{\nu_Og}{\omega_O+g}\right)}

\newcommand{\gs}[1]{{g_{#1}^*}}

\begin{document}

\title{Quasi-steady uptake and bacterial community assembly in a mathematical model of soil-phosphorus mobility}
\author[York]{I.\,R.\ Moyles\corref{cor}}
\ead{imoyles@yorku.ca}

\author[MACSI]{J.\,G.\ Donohue}
\ead{john.donohue.nui@gmail.com}

\author[MACSI,Oxford]{A.\,C.\ Fowler}
\ead{andrew.fowler@ul.ie}

\cortext[cor]{Corresponding Author}

\address[York]{Department of Mathematics and Statistics,
	York University, 4700 Keele Street, Toronto, Ontario, Canada}
\address[MACSI]{MACSI, 
%Department of Mathematics and Statistics, 
University of Limerick, Limerick, Ireland}
\address[Oxford]{OCIAM, 
%Mathematical Institute, 
University of Oxford, 
%24--29 St.~Giles', 
Oxford, UK}

\date{\today}

\begin{abstract}
We mathematically model the uptake of phosphorus by a soil community consisting of a plant and two bacterial groups: copiotrophs and oligotrophs. Four equilibrium states emerge, one for each of the species monopolising the resource and dominating the community and one with coexistence of all species. We show that the dynamics are controlled by the ratio of chemical adsorption to bacterial death permitting either oscillatory states or quasi-steady uptake. We show how a steady state can emerge which has soil and plant nutrient content unresponsive to increased fertilization. However, the additional fertilization supports the copiotrophs leading to community reassembly. Our results demonstrate the importance of time-series measurements in nutrient uptake experiments.
\end{abstract}

\begin{keyword}
Plant-soil (below-ground) interactions \sep nutrient cycling \sep microbial dynamics \sep microbial succession \sep scarce nutrients \sep carbon:phosphorus coupling \sep mathematical ecology
\end{keyword}

\maketitle

\newpage

\section{Introduction}\label{sec:intro}
\begin{figure}[H]
\centering
\includegraphics[scale=0.8]{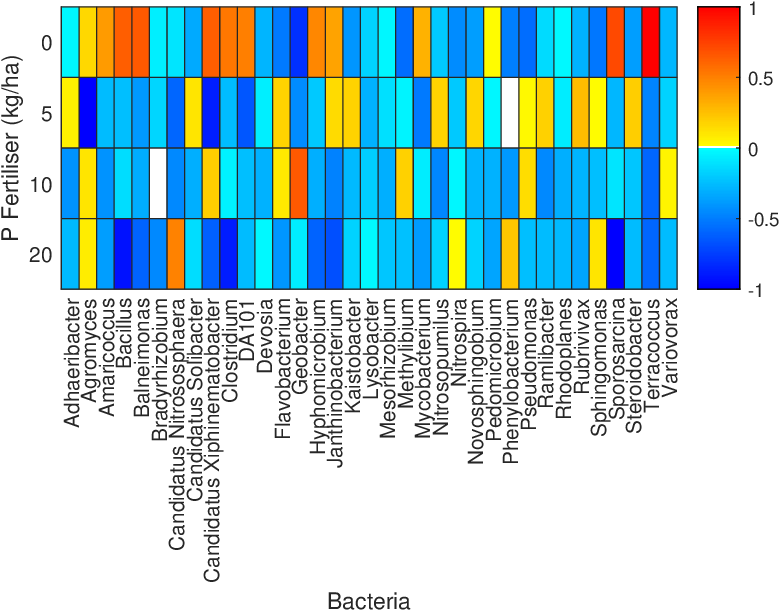}
\caption{Relative bacterial abundance in the soil experiments of \cite{Ikoyi2018}.  Each bacterial species is heated based on its difference from its average abundance and then scaled so that the largest absolute abundance is set to 1 while the lowest to -1.}
\label{fig:heatmap}
\end{figure}

Scarce nutrient supply is one of the greatest problems currently facing humanity.  Nutrients such as carbon, oxygen, and nitrogen, are extremely important for all life on Earth but are readily available from the atmosphere.  Other important nutrients such as phosphorus, sulphur\footnote{Interestingly, during periods of heavy commercial pollution, sulphur was abundant in the atmosphere but levels have since subsided because of stricter environmental regulation (see for example \cite{Baumgardner2002})}, and potassium, must be supplied from external sources to sustain modern agricultural practices.  This problem has been referred to as the broken geochemical cycle by \cite{Elser2011}.  This manuscript will focus on phosphorus since it is the second major growth limiting macronutrient for plants \citep{Schachtman1998, Bergkemper2016}. It plays an important role in processes such as photosynthesis, energy transfer, signal transduction, and legume nitrogen fixation \citep{Kouas2005, Khan2010, Sharma2013}.
\\\\
The environmental flow of phosphorus is unidirectional, starting from rocks where it is mined and ending up in marine ecosystems.  This flow has quadrupled in the last seventy years and \cite{Cordell2009} have predicted that maximum production for phosphorus will occur around 2030 under current uses and operation.  This \textit{phosphate peak forecast} is similar to predictions for peak oil production, however unlike oil and other carbon based products, there is no replacement for phosphorus.  Further complication stems from the localisation of phosphorus to a handful of nations with Morocco identified by \cite{VanKauwenbergh2010} as the country with the most abundant supply.  Overall, these factors contributed to a 700\% increase in the price of rock phosphate from 2007 to 2008 \citep{Elser2011}.  The rapid use and increasing cost of scarce nutrients is related to the difficulties of capturing and recycling the nutrient as well as its inefficient use in soil.  \cite{Cordell2009} studied the fate of about 17.5 million tonnes of mined phosphorus in 2005 where approximately 14 million tonnes of this amount were used in fertiliser.  However, only about 3 million tonnes made it into food products and the rest was lost or stored in the soil. \cite{Khan2009} and \cite{Sharma2013} have estimated that current phosphorus levels in soil could sustain maximal plant yield for 100 years if properly utilised.
\\\\
Phosphorus in soil exists in two forms, organic and inorganic.  Organic forms include soil organic matter or that immobilised into biomass while the inorganic fraction contains precipitates and nutrient adsorbed onto soil mineral surfaces.  \cite{Zou1992} and \cite{Sharma2013} state that only 0.1\% of total phosphorus exists as free orthophosphate, the form required for plant uptake which corresponds to concentrations of 1 ppm or less \citep{Holford1997,Rodriguez1999}.  For plants to access other inorganic forms requires chemical desorption from the soil matrix or solubilisation via the plant or microbial species.  It is clear that greater emphasis must be attached to improving efficiency in fertiliser and the utilisation of nutrients that are already in the soil.
\\\\
Recent experiments by \cite{Ikoyi2018} were designed to elucidate the bacterial transformations of phosphorus in soil.  They set up soil columns (16 $\times$ 40~cm) in a greenhouse environment using a P-limited soil which had not been cultivated for over twenty years.  They planted \emph{Lolium perenne} rye grass and applied single phosphate fertilizer treatments of 0, 5, 10, and 20 kg P ha\unit{-1} alongside a sufficient amount of other nutrients such as nitrogen, potassium, and sulphur.  The columns were monitored for 14 weeks at which point they were deconstructed and plant and root biomass was analysed.  Surprisingly, the phosphate content of the rye grass dry matter yield did not significantly differ between fertilisation levels.  Furthermore, through repeated sampling, Ikoyi \etal also did not observe significant changes in soil phosphorus content. However, they did observe changes in microbial activity and community structure indicating that the fertilisation stimulated biological activity and community reassembly.  The community structure differences are highlighted in Figure \ref{fig:heatmap} which is reproduced from \cite{Ikoyi2018}.  This heat-map shows the relative abundance of various bacterial species found in the soil versus phosphate fertilisation.  By comparing the zero and 20 kg P ha\unit{-1} rows of the heat-map, clear bifurcations in abundance can be seen for several bacterial species. This is a type of microbial succession as discussed by \cite{Fierer2010} whereby disturbances in the soil alter microbial communities. A quantitative theoretical basis for the apparent absence of nutrient uptake and community assembly results of \cite{Ikoyi2018} is lacking and mathematical modelling will be used in this manuscript to address this.

\subsection{Review of Nutrient Modelling}
Mathematical modelling of nutrient cycles and biochemical transformation generally has two approaches.  The most common approach uses computer simulations on large-scale models with coupled compartments.  Large-scale models have been extensively used for nitrogen transformation (cf. \cite{Langergraber2009,Maggi2008,Bailey2015,Louca2016}) but are only starting to be developed for phosphorus.  For example \cite{Slomp2007} consider phosphorus transformation in the ocean leading to historical anoxic events while models by \cite{Runyan2012} and \cite{Runyan2013} consider a soil environment.  These often use data analysis for parameter fitting and the goal is to create a computational environment which mimics certain field or laboratory conditions to predict scientific outcomes.  However, these models generally offer little predictive insight as response markers can rarely be attributed to specific mechanisms.  A contrasting modelling approach is to begin with a reduced model which should capture the crucial behaviour of the system.  While this is a simplification of the problem when compared to the large-scale simulations, an advantage of this approach is that important process mechanisms can easily be isolated and their role on the larger biochemical system evaluated.  The reduced model approach has had success in explaining carbon oscillations \citep{Fowler2014,Mcguinness2014} and nitrate spike formation \citep{Moyles2018} in a contaminated borehole data set.
\\\\
A further assumption of the large-scale models by \cite{Runyan2012} and \cite{Runyan2013} is that the microbial biomass is a single pool contributing to phosphorus cycling.  This ignores changes in nutrient fluxes and chemical pathways in soil because of variation in nutrient concentration.  For example, experiments have demonstrated that the phosphate-specific transport system is repressed by excess phosphate concentration and low external phosphorus concentrations activate the phosphate starvation regulon \citep{Rosenberg1977, Voegele1997, Vershinina2002, Bunemann2012}.  A single microbial pool also disallows the possibility for complex community dynamics such as those displayed by Figure \ref{fig:heatmap}.  By virtue of being simpler, a reduced model framework can naturally include multi-species microbial dynamics.  This, coupled with the ability to provide predictive insight into the underlying biological mechanism, makes the reduced model approach an attractive option and is the one which we will use.
\\\\
This manuscript is outlined as follows.  In section \ref{sec:model} we design a mathematical model for phosphorus uptake in soil with a plant and two bacterial groups.  The plant and bacteria will compete for phosphorus, however the bacterial uptake rate will be coupled to carbon demand, satisfied by plant root exudate. We determine the steady states of the model in section \ref{sec:results} and showcase the interesting dynamics that occur including a steady state where increased fertilization does not impact plant or soil phosphorus levels. We discuss the implications of our model with experimental observations in section \ref{sec:discussion} with conclusions in \ref{sec:conclusion}.

\section{Mathematical Model}\label{sec:model}
\begin{figure}[H]
\centering
\includegraphics[scale=0.8]{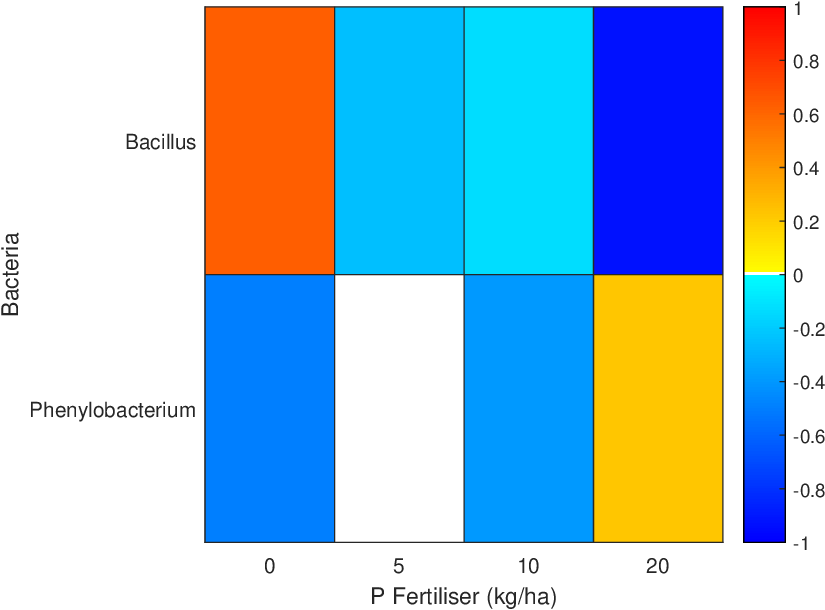}
\caption{Isolation of the species \textit{Bacillus} and \textit{Phenylobacterium} abundance data from Figure \ref{fig:heatmap} which showcases a community structure change from low to high fertilisation.}
\label{fig:heatmap_zoom}
\end{figure}

\begin{figure}[H]
\centering
\includegraphics[scale=0.5]{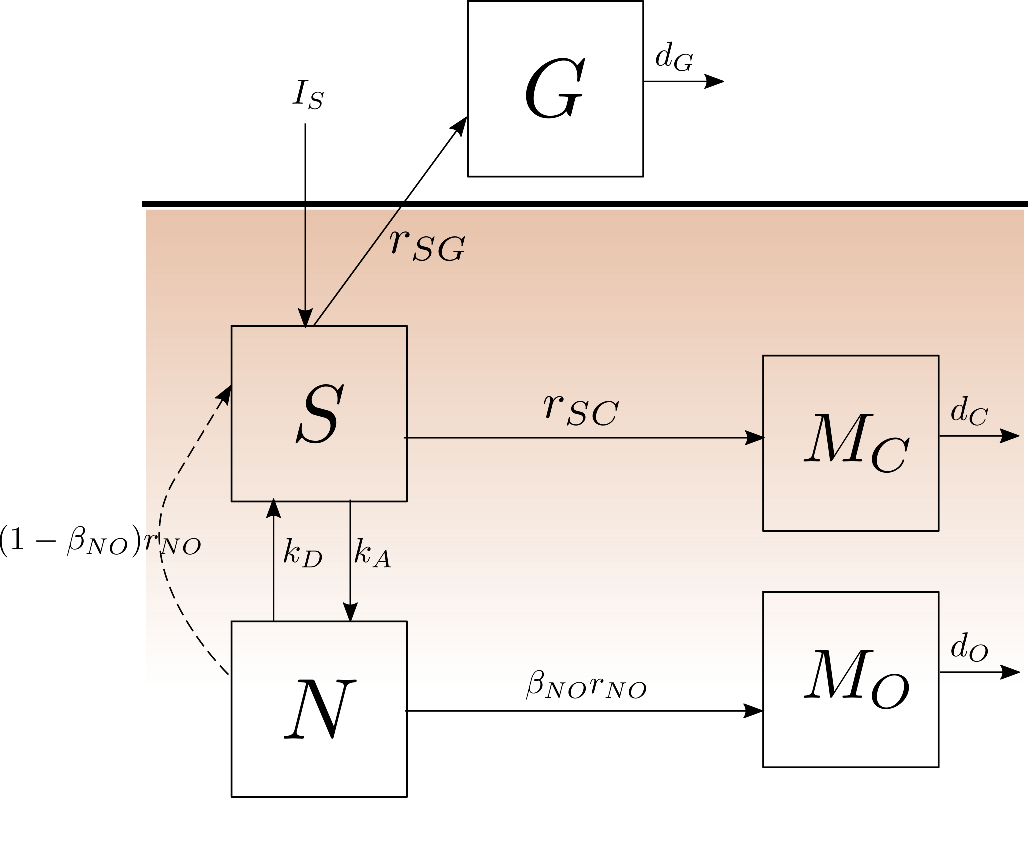}
\caption{Nutrient transformation model in a plant-microbe soil system.  The dashed line indicates that the transformation is a chemical one due to the action of organic acids but it only occurs through biological mediation from the oligotroph.}
\label{fig:model}
\end{figure}

We will consider the nutrient form of phosphorus to exist in two pools, an inorganic dissolved pool with free orthophosphate ($S$) and a slightly occluded pool of adsorbed inorganic phosphorus ($N$) both with units of concentration in milligrams phosphorus per litre soil (mgP L\unit{-1}).  For simplicity we ignore a precipitate pool, however its effects could be included through faster adsorption or slower desorption rates. We also ignore stronger occlusion pools on the notion that this generally results from nutrient ageing and should play a minor role in a short-term fertilized experiment.  An explicit nutrient organic phosphorus pool will not be modelled here because the experiments of \cite{Ikoyi2018} consider inorganic nutrient fertilisation on an initially uncultivated soil.  We therefore assume that the phosphorus response can be accounted for solely by the inorganic transformations.  In agricultural or otherwise cultivated soils with multiple human and natural sources and sinks of phosphorus, an organic pool can be important for nutrient recycling among other functions.
\\\\
We will model the amount of phosphorus stored as plant and bacterial biomass. This is an organic type, but is distinguished from an organic supply pool in that when bacteria and plants die we will consider their phosphorus lost to the system. A mathematical model involving even a modest fraction of the bacterial species detected in the soil experiments of \cite{Ikoyi2018} and represented in Figure \ref{fig:heatmap} would quickly become unmanageable in terms of understanding soil phosphorus mechanisms.  Furthermore, some of the relative abundance changes in Figure \ref{fig:heatmap} were noted as not being statistically significant, and therefore the data could not inform any modelling of such species.  Rather than include a plethora of bacterial species, a more important feature of a phosphorus transformation model is the capture of overall bacterial function.  Several species of varied abundances can exist in soil which all maintain the same function.  This is known as the principle of functional redundancy \citep{Allison2008} and contributes to the mitigation of negative effects from disturbances to soil equilibrium.
\\\\
We will consider a rye-grass plant (as was used by \cite{Ikoyi2018}) which grows via uptake of free orthophosphate $S$ and releases phosphorus through natural decay processes. We will not include the ability for the plant to feed on the inorganic pool, $N$. Although plant roots are able to exude organic acids which allows them to convert inorganic phosphorus to a soluble bioavailable form, it was noted by \cite{Krishnaraj2014} that the highest period of phosphorus demand occurs during juvenile plant development when accessing occluded forms of the nutrient may be difficult, suggesting the need for bacterial support. Furthermore, they also note the other beneficial function that bacteria can have for the plant such as the production of plant growth hormones and antibiotics. For these reasons, we consider the plant focusing its energy on the $S$ pool.
\\\\
There are two primary functions we wish to consider in the bacterial populations.  Firstly, many bacteria will compete directly with the plant for access to the free orthophosphate pool so that they can immobilise phosphorus.  Secondly, some bacteria known as phosphate solubilising bacteria (PSB), can degrade insoluble inorganic phosphorus to forms accessible by plants and other microbes \citep{Chen2006}.  These bacteria will feed on the $N$ pool.  The bacteria \textit{Bacillus} is a known PSB \citep{Kang2014,Swain2012} which was found to have statistically significant variation in relative abundance in \cite{Ikoyi2018}.  Another bacterium which had statistically significant variation in abundance was \textit{Phenylobacterium} which is not known for high phosphate solubilizing activity.  Therefore, these will be the two representative species for the behaviour we wish to model.  A reduced abundance graph isolating these two species is in Figure \ref{fig:heatmap_zoom} where we indeed see a change in community structure from one species to the other while fertilisation rates are varied.
While we consider \textit{Bacillus} and \textit{Phenylobacterium} to be representative species of the important phosphorus activity, in order to generalise the model we will consider two functional bacterial populations, \textit{copiotrophs} and \textit{oligotrophs} as bacteria feeding on $S$ and $N$ respectively.  Copiotrophic bacteria thrive in a high nutrient environment while oligotrophic bacteria have a suppressed metabolism promoting growth in low nutrient supply \citep{Ramirez2012, Pan2014, Bergkemper2016}.  Typically, these terms are reserved for species competing for the same nutrient supply while in this model we use the terms for different forms of the nutrient.  We anticipate that under low supply of free orthophosphate, the PSB thrives due to its ability to utilise adsorbed nutrient. However, under high free orthophosphate supply, the PSB will generally be limited by the amount of nutrient occluded in the adsorbed pool while the grass and plant-competing phosphate bacteria thrive.  Therefore, the resource limitation conditions mimic that of copiotrophic and oligotrophic competition and we use this as a justification for our terminology.   \cite{Ramirez2012} favours this use of a broad classification to better categorise and understand the ecological roles of bacteria.
\\\\
Overall, along with the nutrient pools, $S$, and $N$ we consider three biomass pools, $G$, $M_C$, and $M_O$ for the rye-grass, copiotrophic microbes, and oligotrophic microbes respectively.  We will consider the biomass measurements as the amount of phosphorus stored in milligrams per litre soil (mgP L\unit{-1}).  A pictorial representation of the model is presented in Figure \ref{fig:model} with arrows indicating transformations.  The rates $r_{ij}$ with $i$ as $S$ or $N$ and $j$ as $G$, $C$, or $O$ (mgP L\unit{-1} d\unit{-1}) represent nutrient uptake while $d_i$ (d\unit{-1}) represents loss of organic phosphorus from the system (death).  The rates $k_A$ and $k_D$ (d\unit{-1}) are the adsorption and desorption rates respectively and the dimensionless coefficient $\beta_{ij}\in[0,1]$ is an immobilisation factor and represents the proportion of solubilised $N$ that is taken in by the microbe with the remainder going into the $S$ pool.  This factor exists because solubilising microbes do not uptake $N$ directly but rather convert it to orthophosphate with organic acids which the microorganisms produce \citep{Rodriguez1999}.  Therefore, other organisms have a chance to compete for this newly solubilised orthophosphate.  $I_S$ (mgP L\unit{-1} d\unit{-1}) represents nutrient fertilisation and is taken as being continuous in time. 
\\\\
The fertiliser term includes experimentally applied phosphorus, but could also act as a proxy for natural sources such as nutrient recycling in the absence of an explicit organic pool.  We do not explicitly model leaching because it did not occur in the experiments of \cite{Ikoyi2018}. Furthermore, its inclusion can generally be considered as a reduced fertilisation rate and is therefore in principle already captured in the model.  An initial discrete applied fertilisation event is assumed to be equivalent to a slower continuous fertilisation rate depositing the same amount of phosphorus overall throughout the experiment.  This assumption will almost certainly be invalid in a field-scale or other nutrient system where permanent nutrient losses are significant and nutrient retention times need to be considered.  It also does not include the biological response to a sudden change in nutrient supply. A more general framework of fertiliser forms an avenue for future work.
The experiments of \cite{Ikoyi2018} over-fertilised the other primary nutrients (nitrogen, potassium, sulphur) and therefore we assume that they are sufficiently concentrated that they do not affect the dynamics of the system.  However, carbon was not explicitly added to the system and so we will consider it in our model.  We assume that the rye-grass is not limited by carbon due to its use of photosynthetic processes with carbon dioxide.  However, microbial dynamics will be affected by carbon supply.  \cite{Richardson2011} estimate that up to 20\% of the carbon that plants take up enters the soil as exudate providing a source for micro-organisms.  Rather than model carbon explicitly, we will relate the nutrients through the Redfield ratio.  This is a relatively fixed ratio between carbon and other nutrients first discovered by \cite{Redfield1934} while studying organic matter in plankton.  While the initial ratios concerned ocean nutrient levels, similar ratios have been found to exist in forests \citep{Reich2004,Mcgroddy2004,Hedin2004}, soil microbes \citep{Cleveland2007}, and even entire rhizosphere biomes \citep{Bell2014}.  Values for different plant species in different environments can vary dramatically, however \cite{Bell2014} considered two grass species and reported that the Redfield ratio of the leaves was approximately C:P=$400:1$.  They also reported that carbon levels in roots and leaves were approximately the same, so the Redfield ratio will be sufficient for root exudate.  If we denote the Redfied ratio as $\rho$ then $\rho G$ is approximately the amount of carbon in the plant.  Since the microbes require carbon supplied from the plant we will model their phosphorus uptake to depend on $\rho G$. 

\subsection{Model Equations}
We posit the following mathematical equations for the model presented in Figure \ref{fig:model},
\begin{align}
\begin{split}
\deriv{S}{t}=&I_S-k_AS+k_DN-\frac{r_{SG}}{Y_{SG}}-\frac{r_{SC}}{Y_{SC}}+\frac{(\bbetaNO-\betaNO)}{Y_{NO}}r_{NO},\\
\deriv{N}{t}=&k_AS-k_DN-\frac{\bbetaNO}{Y_{NO}}r_{NO},\\
\deriv{G}{t}=&r_{SG}-d_GG,\\
\deriv{M_C}{t}=&r_{SC}-d_CM_C,\\
\deriv{M_O}{t}=&\betaNO r_{NO}-d_OM_O,
\end{split}\label{eqn:model}
\end{align}
where $Y_{ij}$ are yield coefficients of the organisms utilising the substrate and $\bbeta_{ij}=\beta_{ij}+(1-\beta_{ij})Y_{ij}$ is the effective immobilisation efficiency taking yield into account.  To complete the model, we take the reaction rates in \eqref{eqn:model} to be of Michaelis--Menten (or Monod) form \citep{Johnson2011, Monod1949} which for the plant is simply,
\begin{align}
r_{SG}=\frac{\mu_{SG}SG}{S+k_{SG}}.\label{eqn:rplant}
\end{align}
However, for the microbes we need to incorporate the carbon demand from root exudate.  We will therefore take
\begin{align}
\begin{aligned}
r_{SC}=&\mu_{SC}\left(\frac{SM_C}{S+k_{SC}}\right)\left(1+\nu_C\frac{\rho G}{\rho G+k_{GC}}\right),\\
r_{NO}=&\mu_{NO}\left(\frac{NM_O}{N+k_{NO}}\right)\left(1+\nu_O\frac{\rho G}{\rho G+k_{GO}}\right),
\end{aligned}\label{eqn:rbact}
\end{align}
where $\mu_{ij}$ (d\unit{-1}) are reaction rates for organism $j$ feeding on substrate $i$ with saturation constant $k_{ij}$ (mgP L\unit{-1}).    The parameter $\nu_j$ (dimensionless) is the relative increase in phosphorus uptake due to carbon supply compared to the basal uptake rate of phosphorus in low carbon environments.  If no plant is present ($G=0$) then the microbial uptake rate is $\mu_{ij}$ which is non-zero.  Microbial communities still exist in uncultivated soils, surviving on residual carbon from recycled organic matter and having non-zero $\mu_{ij}$ when $G\equiv0$ represents this effect.  With sufficient plant biomass, the uptake rate of the microbes reaches its maximum boosted rate $\mu_{ij}(1+\nu_j)$.  Multiplicative Monod effects have been considered by other models such as \cite{Blagodatsky1998} and \cite{Moyles2018} and are in contrast to Liebig's law of the minimum used by \cite{Fontaine2005} for example which uses a single Monod term depending on the limiting nutrient.
\subsection{Non-dimensionalisation and Model Reduction}\label{sec:reduce}
We consider scales of the form
\begin{align}
t=t_0\bar{t},\qquad S=S_0s,\qquad N=N_0n, \qquad G=G_0g,\qquad M_C=M_{C0}m_c,\qquad M_O=M_{O0}m_o,
\end{align}
which we substitute into \eqref{eqn:model} to get
\begin{align}
\begin{split}
\dot{s}=&t_ok_A\left(\frac{I_s}{k_AS_0}-s+\frac{k_DN_0}{k_AS_0}n-\left(\frac{\mu_{SG}G_0}{k_{SG}k_AY_{SG}}\right)\rSG\right.\\
&\left.-\left(\frac{\mu_{SC}M_{C0}}{k_{SC}k_AY_{SC}}\right)\ccarb\rSC\right.\\
&\left.+(\bbetaNO-\betaNO)\left(\frac{\mu_{NO}N_0M_{O0}}{k_{NO}k_AS_0Y_{NO}}\right)\ocarb\rNO\right),\\
\dot{n}=&t_0k_A\frac{S_0}{N_0}\left(s-\frac{k_DN_0}{k_AS_0}n-\bbetaNO\left(\frac{\mu_{NO}N_0M_{O0}}{k_{NO}k_AS_0Y_{NO}}\right)\ocarb\rNO\right),\\
\frac{1}{t_0d_G}\dot{g}=&\left(\frac{\mu_{SG}S_0}{k_{SG}d_G}\right)\rSG-g,\\
\frac{1}{t_0d_C}\dot{m}_c=&\left(\frac{\mu_{SC}S_0}{k_{SC}d_C}\right)\ccarb\rSC-m_c,\\
\frac{1}{t_0d_O}\dot{m}_o=&\betaNO\left(\frac{\mu_{NO}N_0}{k_{NO}d_O}\right)\ocarb\rNO-m_o,
\end{split}\label{eqn:model2}
\end{align}
where we have used the dot to indicate differentiation with respect to the non-dimensional time variable $\bar{t}$ and the non-dimensional parameters $\kappa_{ij}$ and $\omega_j$ are defined as
\begin{align}
\kappa_{Sj}=\frac{S_0}{k_{Sj}},\qquad \kappa_{Nj}=\frac{N_0}{k_{Nj}},\qquad \omega_j=\frac{k_{Gj}}{\rho G_0}.
\end{align}
Natural time scale parameters emerge in \eqref{eqn:model2}, which are related to the adsorption chemistry ($t_0k_A$) and to the phosphorus loss of each of the organisms $(t_0d_i)^{-1}$.  We assume that all of these scales are important in the evolution of the system so we define
\begin{align}
t_0k_A=\frac{1}{t_0d_C}:=\delta;\qquad t_0=\sqrt{\frac{1}{k_Ad_C}},\quad \delta=\sqrt{\frac{k_A}{d_C}},\label{eqn:t0}
\end{align}
where we have arbitrarily chosen the copiotrophic death rate as the dominant decay time scale.  Balancing these time scales has led to rich dynamics being captured in other systems such as carbon competition in \cite{Fowler2014} and \cite{Mcguinness2014} as well as for a model of nitrification in soil by \cite{Moyles2018}.  The free orthophosphate scale $S_0$ is most naturally chosen from either the plant or the copiotroph and we will take the latter consistent with choosing $d_C$ as the relevant death rate in \eqref{eqn:t0}.  Therefore,
\begin{align}
S_0=\frac{k_{SC}d_C}{\mu_{SC}}.
\end{align}
The occluded nutrient scale $N_0$ is intimately linked to both uptake by microbes and the adsorption chemistry of the soil.  Since we are particularly interested in solubilisation effects, microbial uptake should be considered the more important scale.  Therefore, we will take
\begin{align}
N_0=\frac{k_{NO}d_O}{\mu_{NO}}.
\end{align}
We anticipate that the dominant sink terms of the soluble phosphate should be the uptake by $g$ and $m_c$ while the dominant uptake of the occluded form should be uptake by $m_o$. This leads to the natural bacterial scales,
\begin{align}
G_0=\frac{Y_{SG}k_{SG}k_A}{\mu_{SG}},\quad M_{C0}=\frac{Y_{SC}k_{SC}k_A}{\mu_{SC}},\quad M_{O0}=\frac{Y_{NO}k_{SC}k_Ad_C}{d_O\mu_{SC}},
\end{align}
each of which come from setting the biological sink terms in \eqref{eqn:model2}\sub{1} to unity.  This scaling assigns equal relevance to adsorption chemistry and microbial uptake which is consistent with the conclusions of \cite{Bunemann2012}.  Finally, we define the following parameters,
\begin{align}
\lambda_G=\frac{d_C}{d_G},\quad \lambda_O=\frac{d_C}{d_O},&\quad \alpha=\frac{k_D}{k_A},\\
\Gamma_I=\frac{I_S}{k_AS_0},\quad \gamma_G=\frac{\mu_{SG}k_{SC}d_C}{\mu_{SC}k_{SG}d_G}&,\quad \gamma_O=\frac{\mu_{NO}k_{SC}d_C}{\mu_{SC}k_{NO}d_O},
\end{align}
to produce the non-dimensional model,
\begin{align}
\begin{split}
\dot{s}=&\delta\left(\Gamma_I-s+\frac{\alpha}{\gamma_O}n-\rSG-\ccarb\rSC\right.\\
&\left.+(\bbetaNO-\betaNO)\ocarb\rNO\right),\\
\dot{n}=&\delta\gamma_O\left(s-\frac{\alpha}{\gamma_O}n-\bbetaNO\ocarb\rNO\right),\\
\delta\lambda_G\dot{g}=&\gamma_G\rSG-g\\
\delta\dot{m}_c=&\ccarb\rSC-m_c\\
\delta\lambda_O\dot{m}_o=&\betaNO\ocarb\rNO-m_o.
\end{split}\label{eqn:nd_model}
\end{align}
We note that while the model \eqref{eqn:nd_model} is motivated by the experimental work in \cite{Ikoyi2018}, it should apply to any system where inorganic nutrient fertilizer is the primary supply of phosphorus. Furthermore, it can be adapted to other nutrients which have a similar bio-geochemical cycle.

We will make assumptions regarding parameter sizes in order to reduce the model to a simpler form for analysis.  Firstly, adsorption and desorption are chemical processes which are relatively well understood.  One estimate by \cite{Barber1995} has the adsorption timescale of phosphorus as approximately 10 days and therefore that $k_A\approx0.1$ d\unit{-1}.  A study of soil buffering in Irish soils was conducted by \cite{Jordan2005} and assuming the $k_A$ value of \cite{Barber1995}, an average desorption rate of $k_D=3.3\tten{-5}$ d\unit{-1} can be extracted from their work.  This leads to the conclusion that $\alpha=3.3\tten{-4}\ll1$ and we will therefore neglect all $\alpha$ terms in \eqref{eqn:nd_model}.  This is consistent with the long-time chemical desorption that occurs for phosphorus.
\\\\
The second assumption we will make in \eqref{eqn:nd_model} is that phosphorus is a limiting nutrient.  This is consistent with the experiments of \cite{Ikoyi2018} and also with many agricultural studies \citep{Swain2012,Bunemann2012,Bergkemper2016,Sharma2013,Harrington2001}.  This means that uptake dynamics should be on the linear growth phase of the Monod kinetics \eqref{eqn:rplant} and \eqref{eqn:rbact} which is equivalent to assuming $\kappa_{ij}\ll1$, i.e., that biological phosphorus concentrations are significantly less than their saturation values.  However, we will assume that the microbes are not sufficiently carbon limited so that $\omega_j\approx\mathcal{O}(1)$.  We note that $\omega_j=\rho^{-1}\kappa_{Gj}^{-1}$ and so the assumption on $\omega_j$ $\mathcal{O}(1)$ works so long as $\rho\gg1$ indicating that there is much more carbon than phosphorus consistent with the reported plant Redfield ratio of $400:1$ by \cite{Bell2014}.
\\\\
Finally, we will assume that the parameters $\lambda_i$, $\gamma_i$, and $\nu_i$ are all $\mathcal{O}(1)$ as well meaning that turnover times and uptake kinetics are roughly equivalent between species and all contribute to the model behaviour. These assumptions may seem unrealistic as proliferation rates and lifetime of plants and bacteria can be drastically different. First we emphasize that $d^{-1}$ is the mean phosphorus turnover time not the mean lifetime of the species. It has been observed by \cite{Richardson2011} that bacteria have a mean turnover time of about 100 days and by \cite{Saggar1996} that ryegrass had a mean turnover time of about 330 days leading to $\lambda_G=3.3$, consistent with the order one approximation. The faster proliferation of bacteria to plants may result in an increased uptake $\mu_{SC}$ compared to $\mu_{SG}$. A value for the former is estimated by \cite{Maier2009} to be 1.1 d$^{-1}$, but a value for the plant uptake is harder to obtain. An estimate is given by \cite{Roose2001} of $F=8.72$ mgP L$^{-1}$ cm d$^{-1}$ which accounts for the plant biomass as well as uptake distance. The experiments of \cite{Ikoyi2018} estimated a plant phosphorus biomass of $B=5.35$ mgP L$^{-1}$ and were performed on $H=40$ cm columns. Using these values we can estimate an uptake rate,
\begin{align}
	\mu_{SG}\approx\frac{F}{BH}=4.1\tten{-2}\textrm{d}^{-1},
\end{align}
which is about 30 times less than bacteria. However, the saturation constants are also different with estimates of $k_{SC}=3$mg L$^{-1}$ from \cite{Maier2009} and $k_{SG}=0.18$mg L$^{-1}$ from \cite{Roose2001}. These values combined with the estimated turnover rates allow us to approximate $\gamma_G\approx2.05$, another order one approximation. Therefore, very disparate growth scales can still lead to order one parameters, a sign that the problem has been sensibly scaled.

Since phosphorus is a limiting nutrient and we have measured biomass via phosphorus, we will take the simplifying assumption that the yield coefficients are unity. This has the effect that $\bbetaNO=1$ in \eqref{eqn:nd_model} and is supported experimentally by \cite{Vrede2002} who studied nutrient limitation in aquatic bacterioplankton. They cultured bacteria in carbon, nitrogen, and phosphorus limited media respectively and measured the amount of biomass attributed to each nutrient after a prescribed growing phase. Using their data, the bacteria had a phosphorus yield coefficient as high as 0.97 in the phosphorus limited media compared to a maximal phosphorus yield of approximately 0.29 in the carbon limited media.
\\\\
%The parameters $\Gamma_I$ and $\delta$ will be dynamic representing various effective fertilisation rates and chemical-to-biological dominance, respectively.  
The final reduced model which will be the focus of the remainder of the manuscript is then
\begin{align}
\begin{split}
\dot{s}=&\delta\left(\Gamma_I-s-sg-\ccarb sm_c\right.\\
&\left.+(1-\betaNO)\ocarb nm_o\right),\\
\dot{n}=&\delta\gamma_O\left(s- \ocarb nm_o\right),\\
\delta\lambda_G\dot{g}=&g(\gamma_Gs-1),\\
\delta\dot{m}_c=&m_c\left(\ccarb s-1\right),\\
\delta\lambda_O\dot{m}_o=&m_o\left(\betaNO \ocarb n-1\right).
\end{split}\label{eqn:simple_mod}
\end{align}

\section{Results}\label{sec:results}
Looking for equilibrium values of \eqref{eqn:simple_mod} produces five results in vector form $[s,n,g,m_c,m_o]$ given by
\begin{align}
\begin{split}
E_0:=&\left[\frac{\Gamma_I}{\betaNO},\frac{1}{\betaNO},0,0,\Gamma_I\right],\\
E_1:=&\left[1,\frac{1}{\betaNO},0,\Gamma_I-\betaNO,\betaNO\right],\\
E_2:=&\left[\frac{1}{\gamma_G},\frac{\gs{1}+\omega_O}{\betaNO((1+\nu_O)\gs{1}+\omega_O)},\gs{1},0,\frac{\betaNO}{\gamma_G}\right],\\
E_3:=&\left[\frac{1}{\gamma_G},\frac{\gs{2}+\omega_O}{\betaNO((1+\nu_O)\gs{2}+\omega_O)},\gs{2},\frac{\gs{1}-\gs{2}}{\gamma_G},\frac{\betaNO}{\gamma_G}\right],
%E_3:=&\left[\frac{1}{\gamma_G},\frac{\gs{2}+\omega_O}{\betaNO((1+\nu_O)\gs{2}+\omega_O)},\gs{2},\Gamma_I-\frac{\bbetaNO\gs{2}+\betaNO}{\bbetaNO\gamma_G},\frac{\betaNO}{\gamma_G\bbetaNO}\right],
\end{split}\label{eqn:sstates}
\end{align}
where,
\begin{align}
\qquad \gs{1}=\Gamma_I\gamma_G-\betaNO,\qquad \gs{2}=\frac{(\gamma_G-1)\omega_C}{1+\nu_C-\gamma_G}.\label{eqn:gs_def}
\end{align}
We classify $E_0$ as the insufficient fertiliser state because only the oligotrophic bacteria survives.  Chemical desorption places the added phosphorus into the occluded pool which is only accessible to the oligotroph.  The amount of phosphorus held by these microbes is precisely that supplied through fertilisation.  States $E_1$ and $E_2$ are copiotroph and plant dominant states respectively.  For these states, the oligotrophs reach a steady value independent of the fertilisation rates while soluble phosphorus competition between copiotrophs and plants ultimately causes one of the populations to go extinct.  Finally, $E_3$ is the triple coexistence state where competition of phosphorus is balanced by bacterial carbon demands.
\\\\
The steady states in \eqref{eqn:sstates} must be positive for biological feasibility which places restriction on all of them except for $E_0$.  For $E_1$, feasibility of the state requires
\begin{align}
\Gamma_I-\betaNO>0,\label{eqn:E1pos}
\end{align}
while for $E_2$, feasibility requires
\begin{align}
\gamma_G\Gamma_I-\betaNO>0.\label{eqn:E2pos}
\end{align}
If \eqref{eqn:E1pos} is satisfied and $\gamma_G>1$ then \eqref{eqn:E2pos} is satisfied as well and conversely if \eqref{eqn:E2pos} is satisfied then so to is \eqref{eqn:E1pos} if $\gamma_G<1$.  Therefore we see that feasibility of $E_1$ and $E_2$ are intimately linked.  
The parameter $\gamma_G$ measures uptake efficiency of the plant compared to the copiotroph.  If $\gamma_G<1$ then the bacteria is more efficient while the grass is more efficient for $\gamma_G>1$.  In fact, if $\gamma_G=1$ then the states $E_1$ and $E_2$ are the same in terms of soil nutrient distribution with only the plant or copiotroph extinct respectively.

Two feasibility constraints exist for positivity of state $E_3$, that both plant and copiotroph nutrient levels are non-zero.  Both of these constraints are satisfied under one of two conditions depending on the parameter $\omega_C$. If $\omega_C\geq\betaNO\nu_C$ then the state $E_3$ is positive if
\begin{subequations}
\begin{align}
\Gamma_I>\betaNO;\qquad 1<\gamma_G<\gamma_G^+.\label{eqn:E3pos1}
\end{align}
Alternatively, if $\omega_C<\betaNO\nu_C$ then the state is positive if
\begin{align}
\Gamma_I>\Gamma_I^+;\qquad \max(1,\gamma_G^-)<\gamma_G<\gamma_G^+,\label{eqn:E3pos2}
\end{align}
\end{subequations}
where
\begin{subequations}
\begin{align}
\begin{split}
\gamma_G^\pm=\frac{p\pm\sqrt{p^2-4\Gamma_I q}}{2\Gamma_I};&\\
p=\betaNO+(1+\nu_C)\Gamma_I-\omega_C,&\qquad q=\betaNO(1+\nu_C)-\omega_C,
\end{split}\label{eqn:pqdef}
\end{align}
and
\begin{align}
\Gamma_I^+=\frac{q+\omega_C\nu_C+2\sqrt{\omega_C\nu_Cq}}{(1+\nu_C)^2}\leq\betaNO. \label{eqn:GamI+def}
\end{align}
\end{subequations}
Details of the positivity of the $E_3$ state can be found in \ref{app:gamgp}.  It is interesting to note that if $\nu_C=0$, i.e., there is no carbon boost to the microbial species from plant exudate, then positivity of $E_3$ fails and the three species can no longer coexist.  This is consistent with the competitive exclusion principle \citep{Hardin1960,Grover1997} which states that when species directly compete for limiting nutrients, one will eventually emerge dominant at the expense of extinction for the others.  Also of note is that if $\omega_C=0$ then there is no plant benefit to the microbes and the triple-existence state vanishes where $E_3$ becomes a plant-extinct state similar to $E_1$, but dependent on $\gamma_G$.  If $\omega_C\to\infty$, i.e., the carbon is nutrient-limiting then $\gs{1}-\gs{2}<0$ and the triple-existence state is infeasible.

\begin{figure}[H]
	\centering
	\subfloat[$\delta=10$, $\gamma_G=0.8$, $\Gamma_I=0.1$] {\includegraphics[width=0.35\textwidth]{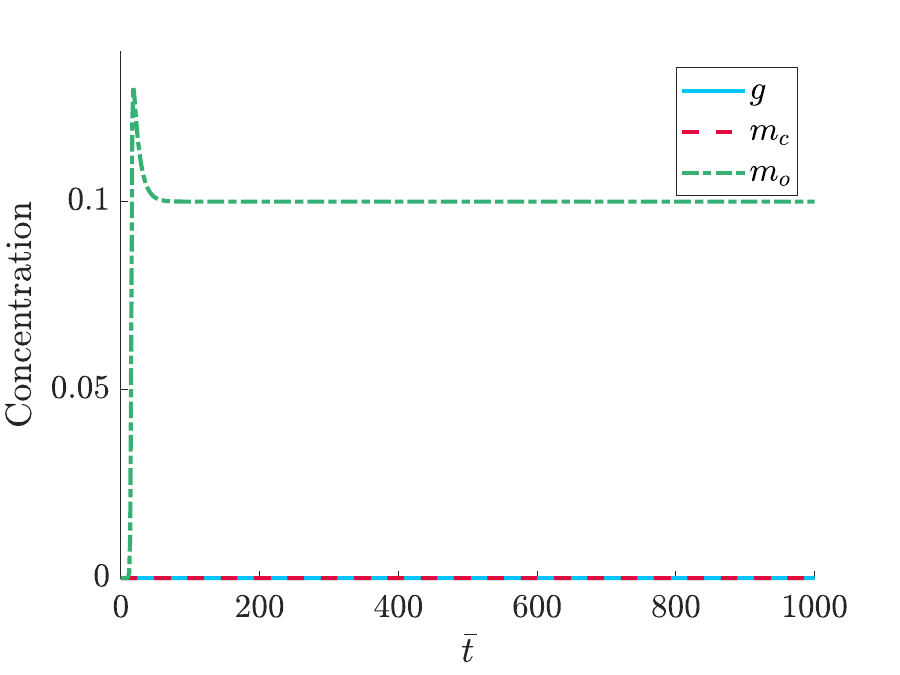}}
	\\
	\subfloat[$\delta=0.07$, $\gamma_G=0.8$, $\Gamma_I=2$]
	{\includegraphics[width=0.35\textwidth]{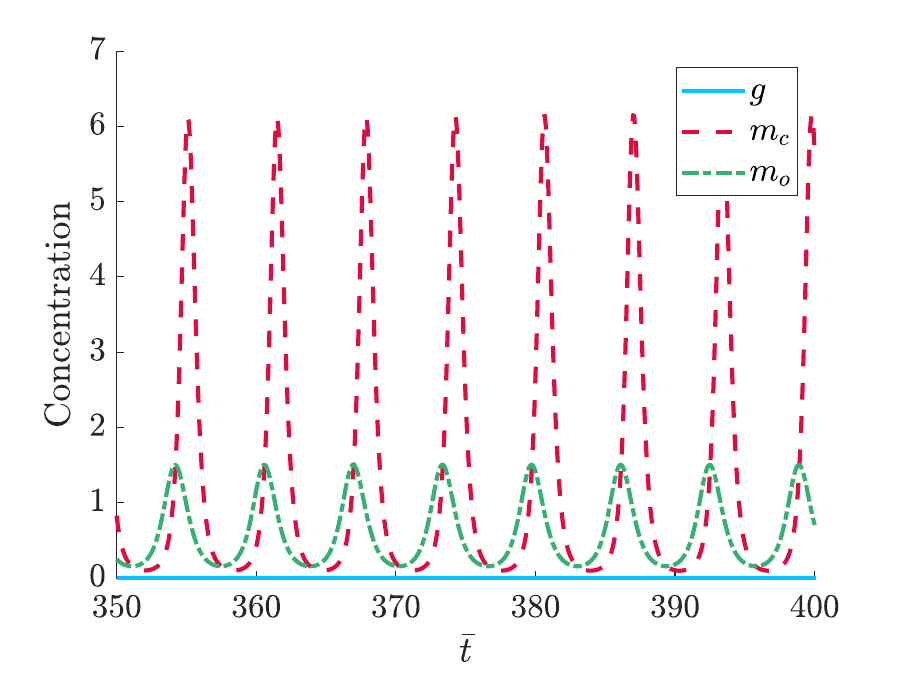}}
	\qquad
	\subfloat[$\delta=10$, $\gamma_G=0.8$, $\Gamma_I=1$]
	{\includegraphics[width=0.35\textwidth]{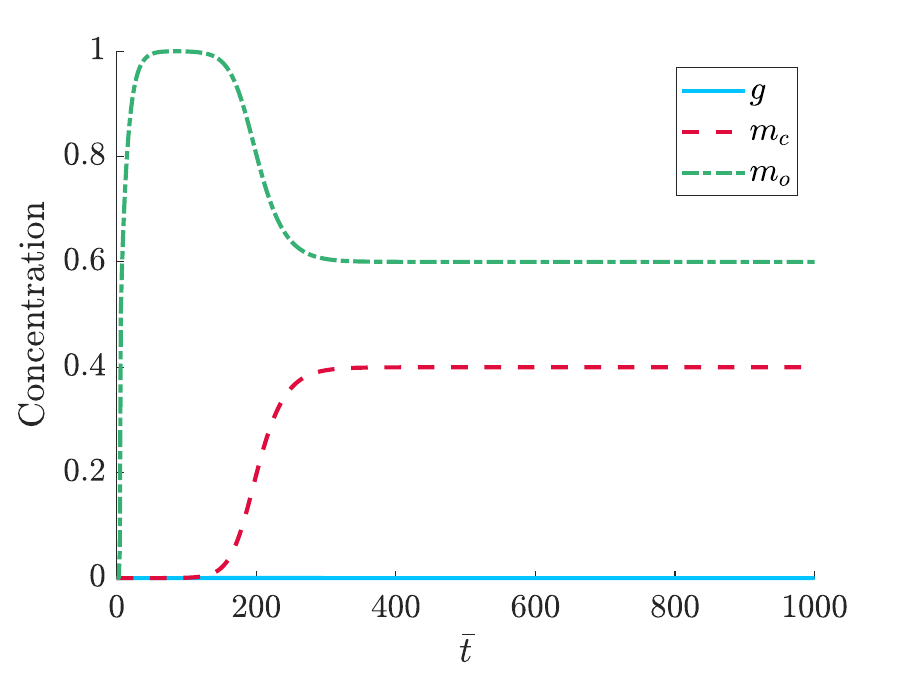}}
	\\
	\subfloat[$\delta=0.1$, $\gamma_G=4$, $\Gamma_I=1$]
	{\includegraphics[width=0.35\textwidth]{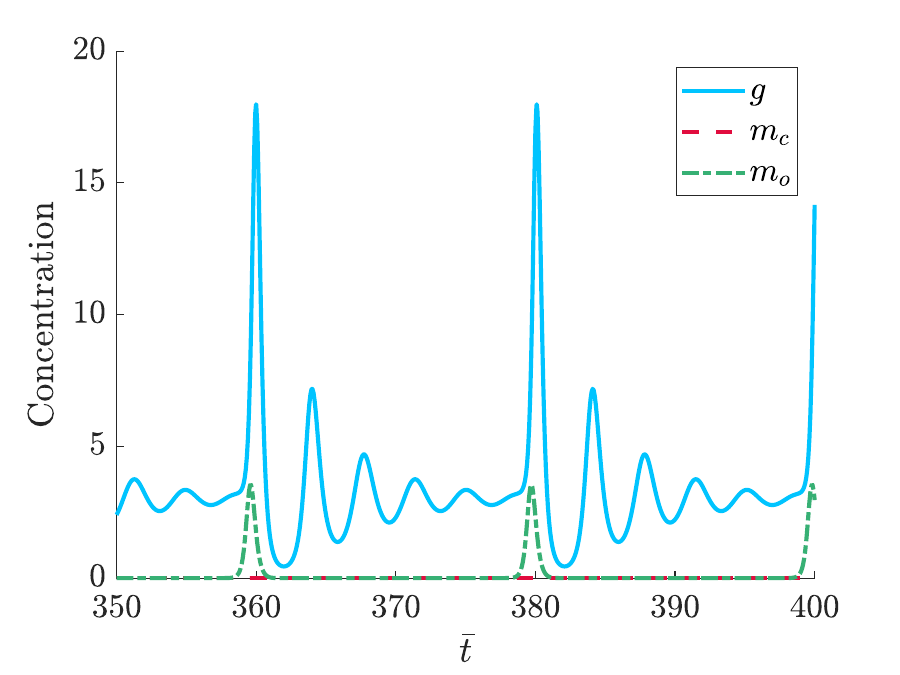}}
	\qquad
	\subfloat[$\delta=10$, $\gamma_G=4$, $\Gamma_I=1$]
	{\includegraphics[width=0.35\textwidth]{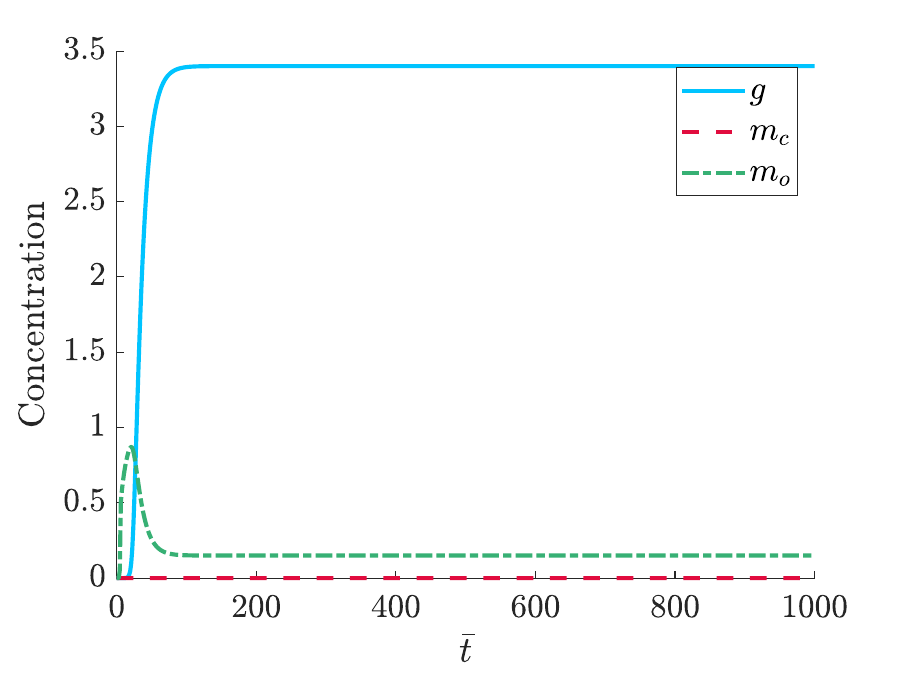}}
	\\
	\subfloat[$\delta=0.1$, $\gamma_G=1.2$, $\Gamma_I=2$]
	{\includegraphics[width=0.35\textwidth]{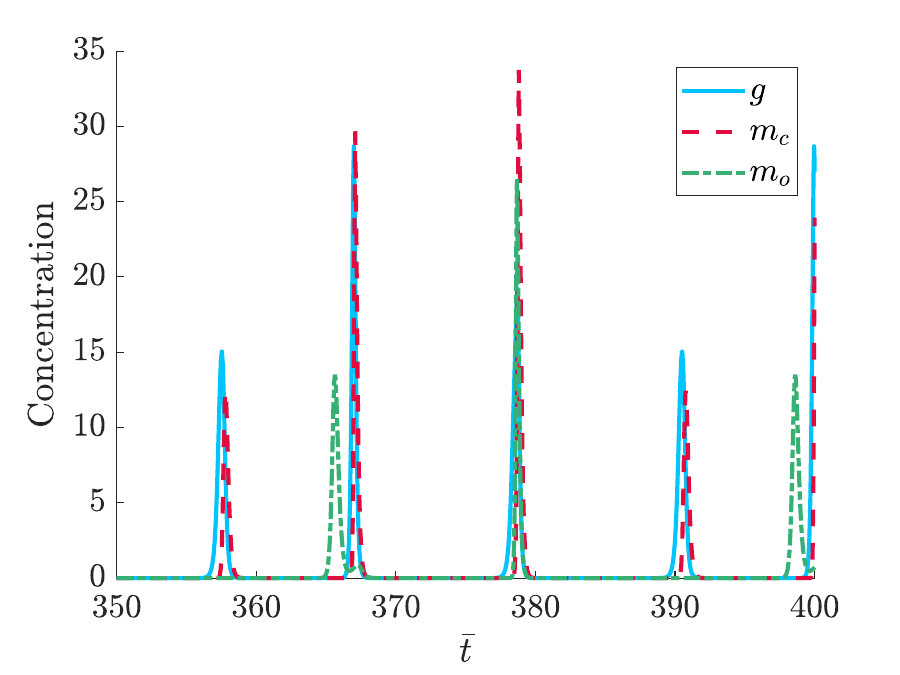}}
	\qquad
	\subfloat[$\delta=10$, $\gamma_G=1.2$, $\Gamma_I=1$]
	{\includegraphics[width=0.35\textwidth]{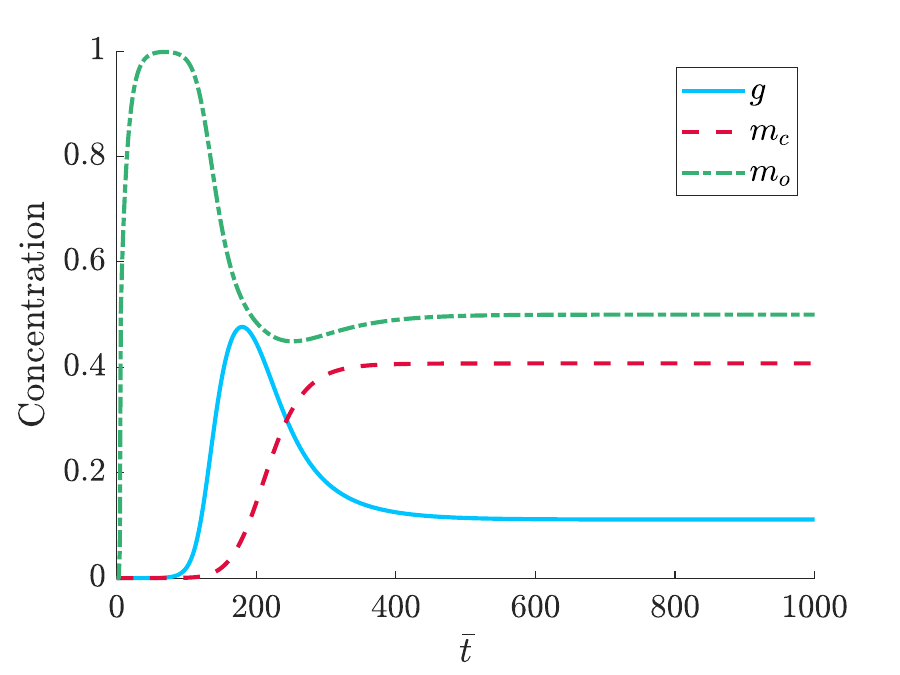}}
	\caption{Plots of plant ($g$), copiotroph ($m_c$), and oligotroph ($m_o$) biomass from numerically solving \eqref{eqn:simple_mod} with fixed parameters $\omega_C=\omega_O=\nu_O=\lambda_O=\lambda_G=1$, $\gamma_O=\nu_C=2$, and $\betaNO=0.6$. The initial conditions are $[s,n,g,m_c,m_o]=[0,0,1\tten{-6},1\tten{-6},1\tten{-6}]$ chosen to avoid extinct populations from the outset. The non-dimensional simulation time is taken as 1000 to allow steady state dynamics (if they occur) to develop. The time window has been reduced in the left panels to highlight the oscillatory dynamics. We omit the plots of $s$ and $n$ for simplicity.}
	\label{fig:dynamics}
\end{figure}

We showcase the rich behaviour of the reduced model in Figure \ref{fig:dynamics} where we numerically solve \eqref{eqn:simple_mod} using {\tt ode45} in MATLAB arbitrarily taking $\omega_C=\omega_O=\nu_O=\lambda_O=\lambda_G=1$, $\gamma_O=\nu_C=2$, and $\betaNO=0.6$ while varying $\Gamma_I$, $\gamma_G$, and $\delta$. We plot only $g$, $m_c$, and $m_o$ for simplicity noting that $s$ and $n$ mimic the behaviour of the other three, i.e., they either always reach a non-zero steady state given by \eqref{eqn:sstates} or oscillate if $g$, $m_c$, and $m_o$ oscillate.

Figure \ref{fig:dynamics} (a), (c), (e), and (g) show the transition through each of the steady states defined by \eqref{eqn:sstates} as $\gamma_G$ and $\Gamma_I$ are varied. Particularly, when $\Gamma_I\ll1$ as in Figure \ref{fig:dynamics} (a) then only the oligotrophs have a non-extinct steady state population. When $\Gamma_I=1>\betaNO$ in Figure \ref{fig:dynamics} (c) the the copiotroph-oligotroph coexistence is the steady state as could be predicted because $\gamma_G<1$ making the carbon state unfeasible. The copiotroph-oligotroph state becomes extinct in favour of the plant-oligotroph state in Figure \ref{fig:dynamics} (e) when $\gamma_G=4>1$. For the fixed parameters in Figure \ref{fig:dynamics}, $\Gamma_I^+=0.38<1$ as defined by \eqref{eqn:GamI+def}. When $\Gamma_I=1$ then $\Gamma_I>\Gamma_I^+$ and $\gamma_G^-=0.36$ and $\gamma_G^+=2.24$ as defined by \eqref{eqn:pqdef}. The triple coexistence state only exists if $1<\gamma_G<2.24<4$ and therefore we could expect that as $\gamma_G$ is reduced below $\gamma_G^+$ that the triple coexistence state will emerge as the equilibrium at the expense of the plant-oligotroph state. This is indeed the case in Figure \ref{fig:dynamics} (g).

Aside from the various steady state profiles expected from \eqref{eqn:sstates}, we also see that oscillatory dynamics can occur in the model as is demonstrated in Figure \ref{fig:dynamics} (b), (d), and (f). These appear to occur when $\delta\ll1$ and oscillatory equivalents exist for each of the states in \eqref{eqn:sstates} except for the insufficient fertilizer state $E_0$. The full classification of the stability of each state \eqref{eqn:sstates} including the emergence of periodic solutions is omitted for brevity and left for future work.

We note that the dynamics are presented in non-dimensional time according to the scaling \eqref{eqn:t0}. Converting to dimensional time requires a knowledge of the dimensional experimental parameters. In the absence of these parameters and inferring values from literature, timescale estimates can vary substantially. For example, following \cite{Richardson2011} bacterial phosphorus turnover occurs every 40-100 days while soil adsorption activity can vary significantly with $k_A=0.1$ d\unit{-1} reported by \cite{Barber1995} while $k_A=48$ d\unit{-1} has been reported from \cite{Santos2011}. These values mean that a non-dimensional time $\bar{t}=1$ following the scaling \eqref{eqn:t0} could correspond to a physical time range of $0.91$ to $31$ days. If we were to use these values for our timescale, the steady states as they appear in Figure \ref{fig:dynamics} would take years or decades to emerge which is unrealistic. We note that the time scale is intimately connected to the value of $\delta$ in \eqref{eqn:t0}. For example taking $d_c^{-1}=100$ d then a 31 day time scale corresponds to $\delta\approx3$ while a 0.91 day time scale would result in $\delta\approx110$. In Figure \ref{fig:dynamics} we arbitrarily take $\delta=10$ when oscillations do not occur to clearly distinguish from the small $\delta$ case where they do. Furthermore, while $\delta$ is related to the timescale, the actual time where the solution has effectively relaxed to equilibrium is intimately dependent on both the initial conditions and the choice of $\gamma_G$. The effect of $\gamma_G$ is evident in Figure \ref{fig:dynamics} (e) and (g) where the plant relaxes to equilibrium around $\bar{t}=100$ for $\gamma_G=4$ and $\bar{t}=400$ for $\gamma_G=1.2$ with all other parameters and initial conditions the same. Overall then, due to a lack of knowledge in experimental parameters to identify an accurate time scale, we present our results in non-dimensional terms.

\begin{figure}[H]
	\centering
	\includegraphics[width=0.65\textwidth]{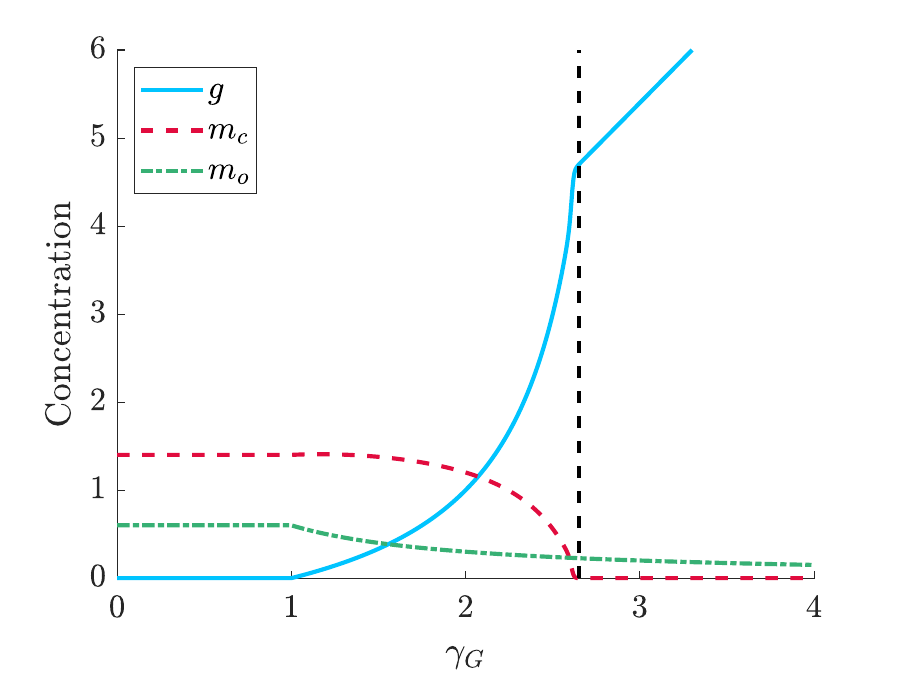}
	\caption{Bifurcation diagram for plants ($g$), copiotrophs ($m_c$), and oligotrophs ($m_o$) via numerically solving the model \eqref{eqn:simple_mod} varying $\gamma_G$ while using parameters $\delta=10$, $\omega_C=\omega_O=\nu_O=\lambda_O=\lambda_G=1$, $\gamma_O=\nu_C=2$, $\betaNO=0.6$, and $\Gamma_I=2$. The initial conditions are $[s,n,g,m_c,m_o]=[0,0,1\tten{-6},1\tten{-6},1\tten{-6}]$, the $s$ and $n$ pools have been omitted, and the steady state is assumed to have occurred by $\bar{t}=5000$.  The vertical black dashed line is the transition value $\gamma_G^+=2.65$ following \eqref{eqn:pqdef} where the triple coexistence state no longer is positive.}
	\label{fig:bif1}
\end{figure}

Due to the absence of oscillatory behaviour in the work of \cite{Ikoyi2018}, we instead focus on the clear bifurcation in steady state solutions as both $\Gamma_I$ and $\gamma_G$ are varied. We illustrate this by plotting the concentration of $g$, $m_c$, and $m_o$ at $\bar{t}=5000$ as $\gamma_G$ varies with $\Gamma_I=2$ fixed (Figure \ref{fig:bif1}), where we see the plant concentration becomes positive as $\gamma_G$ passes through one and the triple coexistence fails for $\gamma_G>\gamma_G^+$. Each of the steady state configurations of Figure \ref{fig:bif1} also appear in the right panels of Figure \ref{fig:dynamics}.

%We note that the final time for computation in Figure \ref{fig:bif1} is taken larger than other simulations in order to guarantee the steady state is reached. 
Observing Figure \ref{fig:dynamics} (c), (e), (g), we note the presence of quasi-steady states that persist for a period of time before relaxing to the final steady state of the system. Taking Figure \ref{fig:dynamics} (g) for example, the $[g,m_c,m_o]$ steady states \eqref{eqn:sstates} are
\begin{align}
\begin{aligned}
E_0=&\left[0,0,1\right],\qquad &E_1=&\left[0,\frac{2}{5},\frac{3}{5}\right]\\
E_2=&\left[\frac{3}{5},0,\frac{1}{2}\right],\qquad &E_3=&\left[\frac{1}{9},\frac{11}{27},\frac{1}{2}\right],
\end{aligned}
\end{align}
and we see that for $\bar{t}<200$ then $E_0$ is the emergent state. There is a brief transition to a plant-oligotroph state with values approaching $E_2$, however the growth of the copiotrophs due to the triple existence quickly sets in and the final state $E_3$ emerges. 

The emergence of quasi-steady states can be attributed to $\delta\gg1$ in \eqref{eqn:simple_mod} where the time derivatives on the populations $g$, $m_c$, and $m_o$ are essentially zero until $\bar{t}\sim\delta$. The precise quasi-steady behaviour that occurs is influenced also by the parameters $\lambda_O$ and $\gamma_G$. In the case of Figure \ref{fig:dynamics} (g), $\lambda_O=\lambda_G=1$ and $\gamma_G>1$ indicating that the plant-dominant state $E_2$ emerges first because $\gamma_G\delta^{-1}>\delta^{-1}$. The quasi-steady state $E_2$ represents the nutrient rich plant and the time frame where it is present represents the ideal agricultural window for harvest. Conversely, $E_3$ can be considered a plant-toxic state in this scenario where the bacteria benefit sufficiently from the carbon exudate of the plant that they are able to overcome their phosphorus uptake deficiency and avail of nutrients that would otherwise be available to the plant. Under these parameters, plant toxicity necessarily occurs because in $E_2$, $g=\gs{1}$ while in $E_3$, $g=\gs{2}$, but necessarily $\gs{1}>\gs{2}$ for $E_3$ to be a feasible state.

\begin{figure}[H]
	\centering
	\subfloat[$\lambda_G=3$, $\gamma_G=1.2$, $\nu_C=2$, $\Gamma_I=1$]
	{\includegraphics[width=0.45\textwidth]{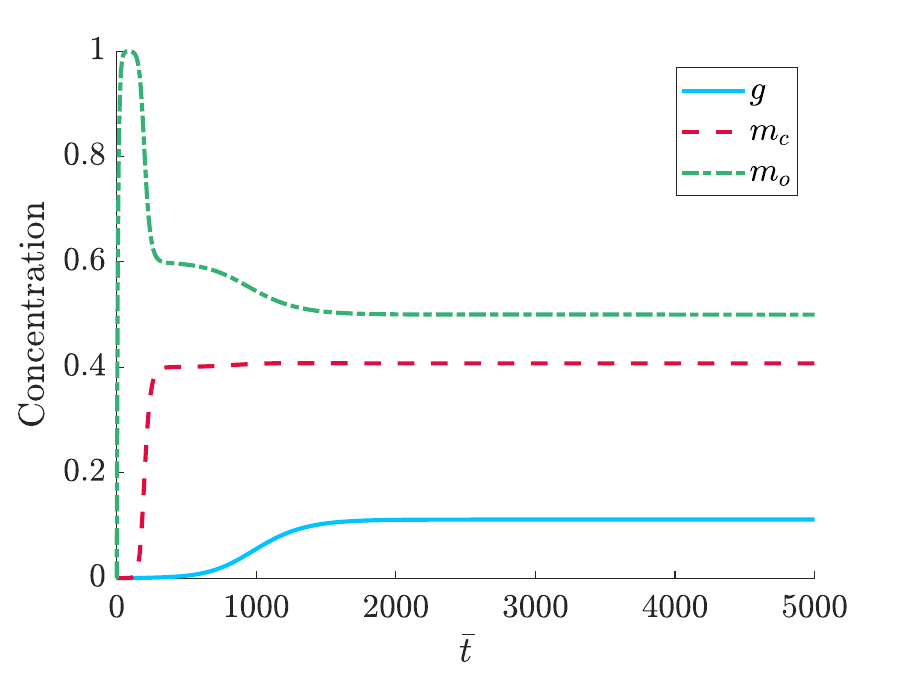}}
	\qquad
	\subfloat[$\lambda_G=3$, $\gamma_G=1.6$, $\nu_C=1$, $\Gamma_I=1.4$]
	{\includegraphics[width=0.45\textwidth]{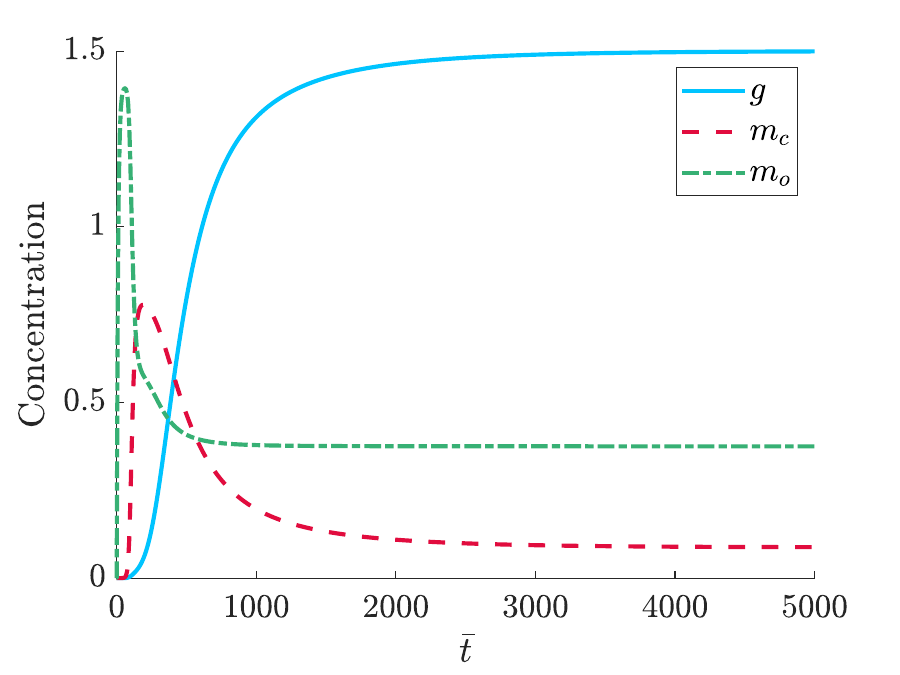}}
	\\
	\subfloat[$\lambda_G=1$, $\gamma_G=2.5$, $\nu_C=2$, $\Gamma_I=2$]
	{\includegraphics[width=0.45\textwidth]{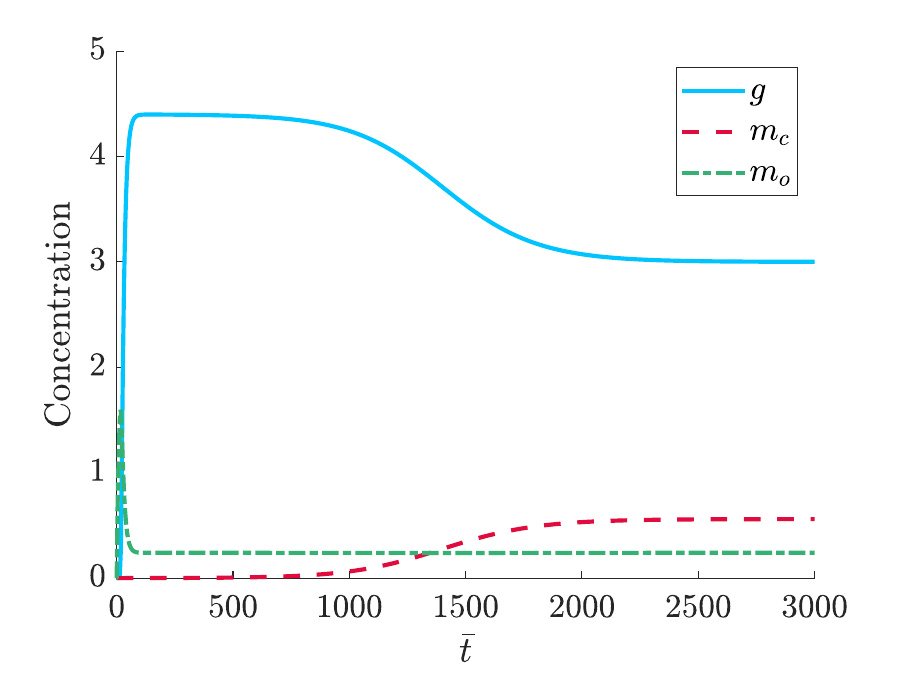}}
	\caption{Simulations of the model \eqref{eqn:simple_mod} varying $\gamma_G$ while using parameters $\delta=10$, $\omega_C=\omega_O=\nu_O=\lambda_O=1$, $\gamma_O=2$, and $\betaNO=0.6$ illustrating non-oscillatory dynamics that do not lead to a plant-toxic scenario. The initial conditions are $[s,n,g,m_c,m_o]=[0,0,1\tten{-6},1\tten{-6},1\tten{-6}]$  The $s$ and $n$ pools have been omitted.}
	\label{fig:otherdynamics}
\end{figure}

For general parameters, it is not necessary that a plant-toxic event occurs which we illustrate in Figure \ref{fig:otherdynamics}. For example, if $\lambda_G>1$ indicating that the nutrient turnover of the plant is slower than that of the copiotrophic bacteria, then if $\gamma_G$ is not very large, the $g$ population remains dormant longer than the copiotroph and state $E_1$ becomes the quasi-steady state before transitioning into the steady state $E_3$ (Figure \ref{fig:otherdynamics} (a)). It is also not necessarily true that $g<m_c$ as the final relative size depends on the fertilization rate, $\Gamma_I$ and $\gamma_G$. An example with $g>m_c$ is in Figure \ref{fig:otherdynamics} (b). Finally, the quasi-steady duration need not be short as is indicated in Figure \ref{fig:otherdynamics} (c). Various parameters can lead to different quasi-steady behaviour for the oligotrophs as well, however this is less important because there is no competitor in the model for the mineralized phosphorus.

Assuming the parameters are such that the biologically feasible triple coexistence state $E_3$ is steady then either $E_2$ or $E_1$ are the early quasi-steady states distinguished by non-zero and zero plant nutrient content respectively. Comparing the plant nutrient content in the states $E_2$ and $E_3$ which are $\gs{1}$ and $\gs{2}$ defined by \eqref{eqn:gs_def} respectively, we see that the former depends on the fertilization level, increasing with further application of nutrient, while the latter is fixed regardless of additional nutrient supply. Therefore, once the state $E_3$ is attained, further nutrient additions only serve to enhance the biomass of the copiotroph microbial community. 

\begin{figure}[H]
	\centering
	\includegraphics[width=0.65\textwidth]{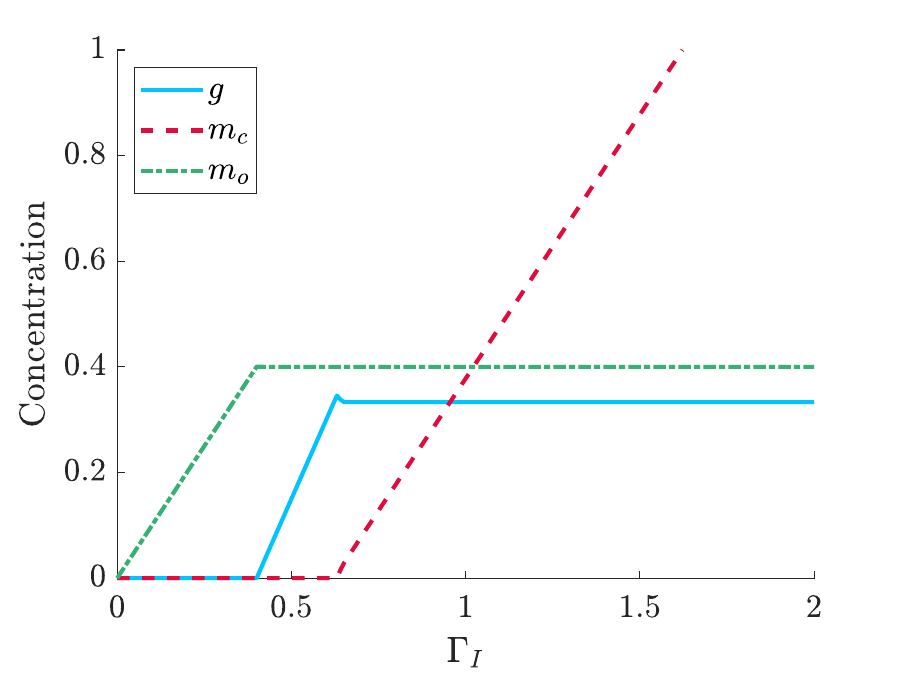}
	\caption{Bifurcation diagram for plants ($g$), copiotrophs ($m_c$), and oligotrophs ($m_o$) via numerically solving the model \eqref{eqn:simple_mod} varying $\Gamma_I$ while using parameters $\omega_C=\omega_O=\nu_O=\lambda_O=\lambda_G=1$, $\delta=10$, $\gamma_O=\nu_C=2$, $\betaNO=0.6$, and $\gamma_G=1.5$. The initial conditions are $[s,n,g,m_c,m_o]=[0,0,1\tten{-6},1\tten{-6},1\tten{-6}]$  The $s$ and $n$ pools have been omitted and the steady state is assumed to have occurred by $\bar{t}=5000$.}
	\label{fig:bif2}
\end{figure}

\begin{figure}[H]
	\centering
	\includegraphics[width=0.65\textwidth]{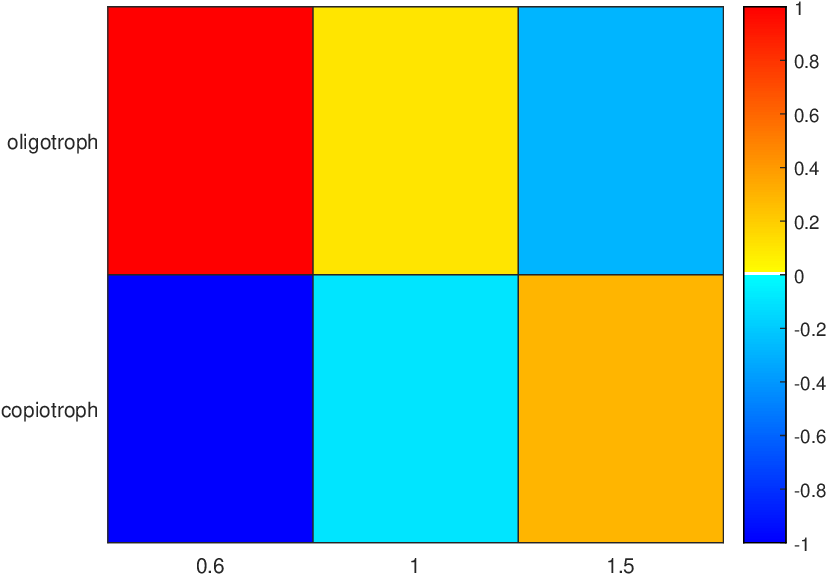}
	\caption{Heat map for copiotrophs ($m_c$) and oligotrophs ($m_o$) via numerically solving the model \eqref{eqn:simple_mod} using the parameters and initial conditions from Figure \ref{fig:bif2} with variable input $\Gamma_I$ indicated on the horizontal axis of the map. The heat map is generated by a relative abundance whereby each bacterial nutrient level is scaled by the total bacterial nutrient level at each input $\Gamma_I$.}
	\label{fig:new_heat}
\end{figure}

While the state $E_3$ permits no benefit to the plant, as additional fertilizer is supplied, the biological dynamics are still quite rich. Figure \ref{fig:bif2} shows a bifurcation diagram where the input, $\Gamma_I$ is varied. We see that for the parameters chosen, the additional fertilizer benefits the copiotrophic bacteria igniting a community reassembly between it and the oligotroph. We also present these results as a heat map in Figure \ref{fig:new_heat} similar to \cite{Ikoyi2018}. To assess relative abundance for this new heat map, we scale each bacterial phosphorus content to the total phosphorus content for a given $\Gamma_I$. As with Figure \ref{fig:heatmap_zoom} we then set the maximum value to $1$ and the minimum to $-1$ to allow for comparison with the experiments of \cite{Ikoyi2018}. We note that there is a good qualitative agreement between Figures \ref{fig:new_heat} and \ref{fig:heatmap_zoom}.

\section{Discussion}\label{sec:discussion}
We have considered a model of nutrient uptake for phosphorus in two inorganic forms: soluble orthophosphate and adsorbed occluded with two bacterial groups and a plant.  We demonstrate that depending on the value of $\delta$, the biological response to nutrient supply is either oscillatory (Figure \ref{fig:dynamics} (b), (d), (f)) or reaches a steady state (Figure \ref{fig:dynamics} (a), (c), (e), (g)).  A spectrum of behaviour has been noted by \cite{Rodriguez1999} and \cite{Babenko1984} where phosphate-solubilizing bacteria were classified into four types based on kinetics and how they accumulated nutrient.  In particular, oscillations are one such category and oscillatory behaviour has been observed by \cite{Illmer1992} among others. During a study of microbial succession, \cite{Fierer2010} noticed oscillatory dynamics in the richness of bacteria taxon.
\\\\
Recently \cite{Ikoyi2018} studied phosphorus uptake in rye-grass and saw very little response in terms of phosphorus content in harvested plant even with high fertilisation levels.  
%While a natural interpretation of this is that phosphorus must not have been the limiting nutrient, Ikoyi \etal debunks this hypothesis as the original soil was phosphorus limited and other nutrients were provided in ample supply along with increasing phosphorus fertilisation.  
They also noticed when sampling the phosphorus content in the soil that levels also remained stagnant with increased fertilization. All of these observations are consistent with the steady state $E_3$ in \eqref{eqn:sstates} of our model \eqref{eqn:simple_mod} where only the copiotrophic nutrient levels depend on the fertilization level with everything else reaching a plateau.
\\\\
Following the nutrient plateau associated with the state $E_3$, one should observe changes in the microbial biomass as fertilisation increases since the copiotroph takes up the additional nutrient.  \cite{Ikoyi2018} measured colonies of bacterial feeding nematodes as a proxy for bacterial populations.  It was observed that during phosphorus fertilisation, the size of nematode colonies decreased except at the highest fertilisation rate (20 kg/ha) where they increased again.  This dynamic nematode response contrasting a static plant response could indicate that the copiotrophs are indeed benefiting from the increased supply of nutrient instead of the plant.  
\\\\
While analysing the model, we have avoided choosing specific parameter values. This is because the necessary parameters are both difficult to find in general and also because they can be quite sensitive to soil and substrate type. Chemical and biological parameters for the soil chosen for the experiments of \cite{Ikoyi2018} were not measured. Considering other literature, most Monod equation kinetic parameters for bacteria have been considered with carbon as a substrate. However, this is beginning to change for two reasons. Primarily, there is renewed focus on phosphorus due to its limited worldwide availability and its limitations on soil activity (see for example \cite{Capek2016,Capek2018}). However, there is also a lot of interest in phosphate solubilizers as bio-fertilizer following, for example, \cite{Meena2016} and \cite{Gumiere2019}, and this requires understanding of the growth kinetics of bacteria such as has been done for \textit{Bacillus} by \cite{Saeid2018}. Plant parameters, particularly ryegrass, have a richer history of study for kinetics and decay such as the works by \cite{Saggar1996,Fohse1988}. However, the uptake parameters of plants can be quite sensitive to root growth which is specific to the growth environment and nutrient supply. Even seemingly static parameters such as soil adsorption and desorption can significantly vary between soil types and agricultural use (see for example \cite{Jordan2005}). We therefore chose to present a general description of the model with arbitrary parameters, showcasing the breadth of behaviour that our simple model can recover. However, we encourage experiments of nutrient dynamics to measure both soil and biota kinetic parameters,  in order to determine which of these behaviours is relevant in a particular setting.
\\\\
Our model predicts a quasi-steady nature to the phosphorus dynamics indicating that sampling should be done frequently to observe various uptake regimes in the system. Depending on the parameters, transitions between states can be quick or very slow. While fast dynamics can lead to a steady state where negligible changes occur as a result of additional fertilization, a consequence of the slow dynamics is that there can be a long time before the system reaches the true equilibrium and an anticipated response observed. This was particularly noted by \cite{Allison2008} who reviewed 110 studies of soil disturbances and observed that timing used for compositional assessment varied from hours to decades. They concluded that studies that reported no effects from disturbances may have seen effects if the study was carried out for longer.  Our model predicts that slow growth depends most strongly on the parameter $\delta$ which can easily be computed by understanding the adsorption time of the dominant nutrient and the turnover time of the copiotroph-like bacteria.  This can therefore be used as a guide in experimental design for setting termination time criteria and when to expect steady state values to emerge.  We have shown how the dimensional equilibration  time is dependent on $\delta$ and $\gamma_G$. A more detailed analysis for large time that furnishes this explicit relationship may help infer values for these parameters by exploiting the experimental duration time as a proxy for equilibration time. However, this would be a coarse approximation and the numerical solutions indicate that time series data as opposed to single point analyses should be considered when reporting results on nutrient dynamics.
\\\\
The bifurcation diagram Figure \ref{fig:bif1} along with the model analysis suggests a natural question to ask would be why a plant would not evolve so that its uptake is sufficient to eliminate the copiotrophic competitor, i.e. to tune the effective $\gamma_G$ parameter so that the $m_c=0$ state is the only stable one?  There are two factors that can prevent this ecological dominance.  Firstly, through evolution and other means such as phenotypic plasticity, the plant can control its uptake parameter but $\gamma_G$ involves the ratio of both plant and copiotroph uptake and its not clear that the species can regulate themselves enough to dramatically affect $\gamma_G$ as the copiotroph will respond by also increasing its uptake.  However, one mechanism that suggests otherwise is that plants have the added support of mycorrhizae, fungal species which use the roots of a plant as a host.  These mycorrhizae use carbon supplied by the plant and in exchange increase the uptake of other nutrients such as phosphorus.  Therefore, it is easy to imagine that the mycorrhizae could in fact promote uptake enough to increase $\gamma_G$.  The second factor, however, that would prevent driving the copiotroph to extinction is that even if the plant can increase the value of $\gamma_G$, the critical point of stability for the coexistence state depends on $\nu_C$ which is the carbon uptake efficiency of the copiotroph and the microbes could counter the plants' attempts to monopolise the nutrient by increasing carbon uptake efficiency, thereby widening the range of coexistence stability.
\\\\
Aside from the mechanisms to control coexistence that we have modelled, there are other factors which promote a symbiotic existence between plant and bacteria.  For example, just as carbon exudate from plants supports bacterial growth, bacterial uptake of nutrient can provide hormonal stimulation to promote root growth in plants as has been observed in \cite{Richardson2009} and \cite{Hayat2010}.  Furthermore, it has been postulated by \cite{Fontaine2003} and \cite{Fontaine2005} that the application of fresh organic matter from plant decay induces a priming effect whereby an increase in native soil organic matter decomposition is observed.  As the microbes mineralise organic forms of the nutrient into more soluble forms for themselves, the plant also benefits by access to that pool.
\\\\
An example which does not rely on balancing effects of cooperative and competitive behaviour is the nutrient recycling provided by microbes.  Microorganisms immobilise nutrients which prevents the plants from taking the nutrient, thus creating a competition.  However, generally the turnover time of bacteria is much smaller than that of plants, so the immobilised nutrient is redistributed into the soil relatively quickly, and following \cite{Macklon1997} and \cite{Oehl2001}, this can lead to measurable increases in soluble concentrations.  It is suggested by \cite{Seeling1993} that immobilisation may be an important mechanism for regulating soluble phosphorus and maintaining it in labile forms available to the plant.  Furthermore, tracer studies show that phosphate released during microbial turnover has a large impact on the basal mineralisation rates in soil (see \cite{Richardson2011}) which could be similar to the priming effect of fresh organic matter discussed by \cite{Fontaine2003}.  It is therefore of interest in future work to consider a model similar to that of \cite{Fontaine2005}, but with the fresh organic matter being supplied by the microbes, thus quantifying the benefits of recycling.

\subsection{Conclusions}\label{sec:conclusion}
Our model provides an explanation for how static soil nutrient levels and uptake in plants can be observed and is a consequence of microbial activity in soil.  We can explain the unintuitive result of high microbe activity occurring simultaneously with low plant activity as seen in experiments such as \cite{Ikoyi2018}.  We have confirmed community shifts between bacterial species and bacteria and plant as nutrient supply is varied which is consistent with results in other literature.  Our model clearly showcases the interesting dynamics that can occur even when steady state behaviour is anticipated and an important implication of this is a greater need for experiments to be designed with time series data in mind.  Time series data can be used to refine this model by assisting with parameter estimation and prediction for the time scale of dominant plant uptake.  Furthermore, the data can better refine the modelling efforts and illuminate processes that are redundant to model or others which are not yet considered.

\subsection*{Acknowledgements}
\noindent This publication has emanated from research conducted with the financial support of Science Foundation Ireland under grant number SFI/13/IA/1923.  I.\,R.\,M.\ acknowledges an NSERC Discovery Grant (2019-06337). A.\,C.\,F.\ acknowledges support from the Mathematics Applications Consortium for Science and Industry ({\tt www.macsi.ul.ie}) funded by the Science Foundation Ireland grant 12/IA/1683. The authors would like to thank Achim Schmalenberger and Israel Ikoyi for valuable discussions about this work.\\

\textbf{Declarations of Interest:} None

\subsection*{Author's Contribution}
All authors jointly devised the paper. I.R.M. constructed the model and wrote the paper. All authors performed the literature review and worked on the final text.\\

\appendix
%%%%%%%%%%%%%%%%%%%%%%%%%%%%%%%%%%%%%%%%%%%%%%%%%%%%%%%%%%%%%%%%%%%%%%%%%%%%%%%%%%%%%%%%%%%%%%%%%%%%%%%%%%%%%%%%%%%%%%%%%%%%%%%%%%%%%%%%%%%%%%%%%%%%%%%%%%%%%%%%%%%%%%%%%%%%%%%%%%%%%%%%%%%%%
\section{Positivity of Copiotroph Bacteria in Coexistence State, $E_3$}\label{app:gamgp}
The coexistence state $E_3$ in \eqref{eqn:sstates} has as a steady state value for the plant, $g=\gs{2}$ and copiotroph, $m_C=\tfrac{\gs{1}-\gs{2}}{\gamma_G}$. Feasibility of $E_3$ requires each to be positive.  First we note that if $\nu_C=0$ then $\gs{2}=-\omega_c\leq0$ and so we will assume $\nu_C>0$. Using the definition of $\gs{1}$ and $\gs{2}$ from \eqref{eqn:gs_def}, $\gs{2}>0$ if $1<\gamma_G<1+\nu_C$, while $\gs{1}-\gs{2}>0$ if
\begin{align}
\begin{split}
h(\gamma_G)=\Gamma_I\gamma_G^2-p\gamma_G+q&<0;\\
p=\betaNO+(1+\nu_C)\Gamma_I-\omega_C,& \qquad q=\betaNO(1+\nu_C)-\omega_C.
\end{split}\label{eqn:hgam}
\end{align}
The roots of this are given by
\begin{align}
\gamma_G^\pm=\frac{p\pm\sqrt{f(\Gamma_I)}}{2\Gamma_I};\qquad f(\Gamma_I)=p^2-4\Gamma_I q,\label{eqn:gamroots}
\end{align}
and since $h''(\gamma_G)>0$ then $h<0$ on $\gamma_G^-<\gamma_G<\gamma_G^+$ if such roots exist. Roots to \eqref{eqn:gamroots} exist provided that $f(\Gamma_I)>0$ which is true on $[0,\Gamma_I^-)\cup(\Gamma_I^+,\infty)$ where,
\begin{align}
\Gamma_I^\pm=\frac{q+\omega_C\nu_C\pm 2\sqrt{\omega_C\nu_Cq}}{(1+\nu_C)^2}.
\end{align}
Roots to this exist for $q\omega_C>0$ which is true for $0\leq q\leq\betaNO(1+\nu_C)$. If $q<0$ then $f>0$ always and $q$ is never greater than $\betaNO(1+\nu_C)$ because $\omega_C\geq0$.

We note a couple of properties of the roots. Firstly, if $\Gamma_I\gg1$ then
\begin{align}
\gamma_G^+\sim 1+\nu_C-\frac{1}{\Gamma_I}\frac{\betaNO(1+\nu_C)-q}{1+\nu_C}\leq 1+\nu_C
\end{align}
since $q\leq \betaNO(1+\nu_C)$. Therefore $\gamma_G^+$ is bounded from above by $1+\nu_C$. Secondly, we note that differentiating $h(\gamma_G^\pm)=0$ with respect to $\Gamma_I$ yields,
\begin{align}
\deriv{\gamma_G^\pm}{\Gamma_I}=\mp\frac{\gamma_G^\pm(\gamma_G^\pm-(1+\nu_C))}{\sqrt{f(\Gamma_I)}},\label{eqn:gamder}
\end{align}
from which we conclude that on $0<\gamma<1+\nu_C$, $\gamma_G^+$ is an increasing function of $\Gamma_I$ while $\gamma_G^-$ is a decreasing function. Thirdly, solving $h(1)=0$ in \eqref{eqn:hgam} yields $\Gamma_I=\betaNO$. Conversely then, if $\Gamma_I=\betaNO$ then upon solving \eqref{eqn:hgam} 
\begin{align}
\gamma_G^-=\min\left(1,\frac{q}{\betaNO}\right),\qquad \gamma_G^+=\max\left(1,\frac{q}{\betaNO}\right).\label{eqn:root_lims}
\end{align}

We are now in a position to classify the roots. From \eqref{eqn:root_lims} we will separately consider $q\leq\betaNO$ and $q>\betaNO$. If $q\leq\betaNO$ and $\Gamma_I=\betaNO$ then from \eqref{eqn:root_lims} $\gamma_G^-=q/\betaNO<1$ and $\gamma_G^+=1$. In light of \eqref{eqn:gamder}, $\gamma_G^+<1$ if $\Gamma_I<\betaNO$ and therefore it is not possible to satisfy $g>0$ and $m_C>0$ in this case. We therefore conclude that for $q\leq\betaNO$, $g$ and $m_C$ are both positive on $1<\gamma_G<\gamma_G^+$ if $\Gamma_I>\betaNO$.

Now consider $q>\betaNO$ where if $\Gamma_I=\betaNO$ then $\gamma_G^-=1$ and $\gamma_G^+=q/\betaNO>1$. If $\Gamma_I>\betaNO$ then $\gamma_G^-<1$ and $\gamma_G^+>q/\betaNO$ so like before we conclude that both $g$ and $m_C$ are positive on $1<\gamma_G<\gamma_G^+$ so long as $\Gamma_I>\betaNO$. However, if $\Gamma_I<\betaNO$ then $\gamma_G^->1$ and our existence interval is reduced. If we differentiate $f(\Gamma_I^\pm)=0$ in \eqref{eqn:gamroots} we discover that $\Gamma_I^-$ has a minimum $\Gamma_I^-=0$ when $q=\nu_C\betaNO$ and $\Gamma_I^+$ has a maximum $\Gamma_I^+=\betaNO$ when $q=\betaNO$. Since $q>\betaNO$ then $\Gamma_I^+<\betaNO$. Furthermore, since $f(\Gamma_I)>0$ on $\Gamma_I^+<\Gamma_I<\infty$ then as $\Gamma_I$ decreases from $\betaNO$ a solution with $g$ and $m_C$ positive exists on $\gamma_G^-<\gamma_G<\gamma_G^+$ for $\Gamma_I^+<\Gamma_I<\betaNO$.

Overall then, there are two positivity conditions on $E_3$. If $1<\gamma_G<\gamma_G^+$ with $\omega_C\geq\betaNO\nu_C$ and $\Gamma_I>\betaNO$ then both $g$ and $m_C$ are positive. Alternatively, these states are positive on $\max(1,\gamma_G^-)<\gamma_G<\gamma_G^+$ if $\omega_C<\betaNO\nu_C$ and $\Gamma_I>\Gamma_I^+$.

%%%%%%%%%%%%%%%%%%%%%%%%%%%%%%%%%%%%%%%%%%%%%%%%%%%%%%%%%%%%%%%%%%%%%%%%%%%%%%%%%%%%%%%%%%%%%%%%%%%%%%%%%%%%%%%%%%%%%%%%%%%%%%%%%%%%%%%%%%%%%%%%%%%%%%%%%%%%%%%%%%%%%%%%%%%%%%%%%%%%%%%%%%%%%

\newpage
%\section*{References}
\bibliographystyle{apalike}
\bibliography{bib(corr_hidden)}

\begin{thebibliography}{}

\bibitem[Allison and Martiny, 2008]{Allison2008}
Allison, S.~D. and Martiny, J.~B. (2008).
\newblock Resistance, resilience, and redundancy in microbial communities.
\newblock {\em Proceedings of the National Academy of Sciences}, 105(Supplement
  1):11512--11519.

\bibitem[Babenko et~al., 1984]{Babenko1984}
Babenko, Y., Tyrygina, G., Grigoryev, E., Dolgikh, L., and Borisova, T. (1984).
\newblock Biological activity and physiologo-biochemical properties of bacteria
  dissolving phosphates.
\newblock {\em Microbiologiya}, 53(4):533--539.

\bibitem[Bailey et~al., 2015]{Bailey2015}
Bailey, R.~T., Gates, T.~K., and Romero, E.~C. (2015).
\newblock Assessing the effectiveness of land and water management practices on
  nonpoint source nitrate levels in an alluvial stream--aquifer system.
\newblock {\em Journal of Contaminant Hydrology}, 179:102--115.

\bibitem[Barber, 1995]{Barber1995}
Barber, S.~A. (1995).
\newblock {\em Soil nutrient bioavailability: a mechanistic approach}.
\newblock John Wiley \& Sons.

\bibitem[Baumgardner et~al., 2002]{Baumgardner2002}
Baumgardner, R.~E., Lavery, T.~F., Rogers, C.~M., and Isil, S.~S. (2002).
\newblock Estimates of the atmospheric deposition of sulfur and nitrogen
  species: Clean air status and trends network, 1990- 2000.
\newblock {\em Environmental Science \& Technology}, 36(12):2614--2629.

\bibitem[Bell et~al., 2014]{Bell2014}
Bell, C., Carrillo, Y., Boot, C.~M., Rocca, J.~D., Pendall, E., and
  Wallenstein, M.~D. (2014).
\newblock Rhizosphere stoichiometry: are {C}: {N}: {P} ratios of plants, soils,
  and enzymes conserved at the plant species-level?
\newblock {\em New Phytologist}, 201(2):505--517.

\bibitem[Bergkemper et~al., 2016]{Bergkemper2016}
Bergkemper, F., Sch{\"o}ler, A., Engel, M., Lang, F., Kr{\"u}ger, J., Schloter,
  M., and Schulz, S. (2016).
\newblock Phosphorus depletion in forest soils shapes bacterial communities
  towards phosphorus recycling systems.
\newblock {\em Environmental Microbiology}, 18(6):1988--2000.

\bibitem[Blagodatsky and Richter, 1998]{Blagodatsky1998}
Blagodatsky, S. and Richter, O. (1998).
\newblock Microbial growth in soil and nitrogen turnover: a theoretical model
  considering the activity state of microorganisms.
\newblock {\em Soil Biology and Biochemistry}, 30(13):1743--1755.

\bibitem[B{\"u}nemann et~al., 2012]{Bunemann2012}
B{\"u}nemann, E., Oberson, A., Liebisch, F., Keller, F., Annaheim, K.,
  Huguenin-Elie, O., and Frossard, E. (2012).
\newblock Rapid microbial phosphorus immobilization dominates gross phosphorus
  fluxes in a grassland soil with low inorganic phosphorus availability.
\newblock {\em Soil Biology and Biochemistry}, 51:84--95.

\bibitem[{\v{C}}apek et~al., 2016]{Capek2016}
{\v{C}}apek, P., Kotas, P., Manzoni, S., and {\v{S}}antr{\r{u}}{\v{c}}kov{\'a},
  H. (2016).
\newblock Drivers of phosphorus limitation across soil microbial communities.
\newblock {\em Functional Ecology}, 30(10):1705--1713.

\bibitem[{\v{C}}apek et~al., 2018]{Capek2018}
{\v{C}}apek, P., Manzoni, S., Ka{\v{s}}tovsk{\'a}, E., Wild, B.,
  Di{\'a}kov{\'a}, K., B{\'a}rta, J., Schnecker, J., Biasi, C., Martikainen,
  P.~J., Alves, R. J.~E., et~al. (2018).
\newblock A plant--microbe interaction framework explaining nutrient effects on
  primary production.
\newblock {\em Nature Ecology \& Evolution}, 2(10):1588.

\bibitem[Chen et~al., 2006]{Chen2006}
Chen, Y., Rekha, P., Arun, A., Shen, F., Lai, W.-A., and Young, C. (2006).
\newblock Phosphate solubilizing bacteria from subtropical soil and their
  tricalcium phosphate solubilizing abilities.
\newblock {\em Applied Soil Ecology}, 34(1):33--41.

\bibitem[Cleveland and Liptzin, 2007]{Cleveland2007}
Cleveland, C.~C. and Liptzin, D. (2007).
\newblock {C}: {N}: {P} stoichiometry in soil: is there a “redfield ratio”
  for the microbial biomass?
\newblock {\em Biogeochemistry}, 85(3):235--252.

\bibitem[Cordell et~al., 2009]{Cordell2009}
Cordell, D., Drangert, J.-O., and White, S. (2009).
\newblock The story of phosphorus: global food security and food for thought.
\newblock {\em Global Environmental Change}, 19(2):292--305.

\bibitem[Elser and Bennett, 2011]{Elser2011}
Elser, J. and Bennett, E. (2011).
\newblock Phosphorus cycle: a broken biogeochemical cycle.
\newblock {\em Nature}, 478(7367):29--31.

\bibitem[Fierer et~al., 2010]{Fierer2010}
Fierer, N., Nemergut, D., Knight, R., and Craine, J.~M. (2010).
\newblock Changes through time: integrating microorganisms into the study of
  succession.
\newblock {\em Research in microbiology}, 161(8):635--642.

\bibitem[F{\"o}hse et~al., 1988]{Fohse1988}
F{\"o}hse, D., Claassen, N., and Jungk, A. (1988).
\newblock Phosphorus efficiency of plants.
\newblock {\em Plant and Soil}, 110(1):101--109.

\bibitem[Fontaine and Barot, 2005]{Fontaine2005}
Fontaine, S. and Barot, S. (2005).
\newblock Size and functional diversity of microbe populations control plant
  persistence and long-term soil carbon accumulation.
\newblock {\em Ecology Letters}, 8(10):1075--1087.

\bibitem[Fontaine et~al., 2003]{Fontaine2003}
Fontaine, S., Mariotti, A., and Abbadie, L. (2003).
\newblock The priming effect of organic matter: a question of microbial
  competition?
\newblock {\em Soil Biology and Biochemistry}, 35(6):837--843.

\bibitem[Fowler et~al., 2014]{Fowler2014}
Fowler, A., Winstanley, H., McGuinness, M., and Cribbin, L. (2014).
\newblock Oscillations in soil bacterial redox reactions.
\newblock {\em Journal of Theoretical Biology}, 342:33--38.

\bibitem[Grover, 1997]{Grover1997}
Grover, J.~P. (1997).
\newblock {\em Resource competition}, volume~19.
\newblock Springer Science \& Business Media.

\bibitem[Gumiere et~al., 2019]{Gumiere2019}
Gumiere, T., Rousseau, A.~N., da~Costa, D.~P., Cassetari, A., Cotta, S.~R.,
  Andreote, F.~D., Gumiere, S.~J., and Pavinato, P.~S. (2019).
\newblock Phosphorus source driving the soil microbial interactions and
  improving sugarcane development.
\newblock {\em Scientific Reports}, 9(1):4400.

\bibitem[Hardin, 1960]{Hardin1960}
Hardin, G. (1960).
\newblock The competitive exclusion principle.
\newblock {\em Science}, 131(3409):1292--1297.

\bibitem[Harrington et~al., 2001]{Harrington2001}
Harrington, R.~A., Fownes, J.~H., and Vitousek, P.~M. (2001).
\newblock Production and resource use efficiencies in {N}-and {P}-limited
  tropical forests: a comparison of responses to long-term fertilization.
\newblock {\em Ecosystems}, 4(7):646--657.

\bibitem[Hayat et~al., 2010]{Hayat2010}
Hayat, R., Ali, S., Amara, U., Khalid, R., and Ahmed, I. (2010).
\newblock Soil beneficial bacteria and their role in plant growth promotion: a
  review.
\newblock {\em Annals of Microbiology}, 60(4):579--598.

\bibitem[Hedin, 2004]{Hedin2004}
Hedin, L.~O. (2004).
\newblock Global organization of terrestrial plant--nutrient interactions.
\newblock {\em Proceedings of the National Academy of Sciences},
  101(30):10849--10850.

\bibitem[Holford, 1997]{Holford1997}
Holford, I. (1997).
\newblock Soil phosphorus: its measurement, and its uptake by plants.
\newblock {\em Soil Research}, 35(2):227--240.

\bibitem[Ikoyi et~al., 2018]{Ikoyi2018}
Ikoyi, I., Fowler, A., and Schmalenberger, A. (2018).
\newblock One-time phosphate fertilizer application to grassland columns
  modifies the soil microbiota and limits its role in ecosystem services.
\newblock {\em Science of The Total Environment}, 630:849--858.

\bibitem[Illmer and Schinner, 1992]{Illmer1992}
Illmer, P. and Schinner, F. (1992).
\newblock Solubilization of inorganic phosphates by microorganisms isolated
  from forest soils.
\newblock {\em Soil Biology and Biochemistry}, 24(4):389--395.

\bibitem[Johnson and Goody, 2011]{Johnson2011}
Johnson, K.~A. and Goody, R.~S. (2011).
\newblock The original {M}ichaelis constant: translation of the 1913
  {M}ichaelis--{M}enten paper.
\newblock {\em Biochemistry}, 50(39):8264--8269.

\bibitem[Jordan et~al., 2005]{Jordan2005}
Jordan, P., Menary, W., Daly, K., Kiely, G., Morgan, G., Byrne, P., and Moles,
  R. (2005).
\newblock Patterns and processes of phosphorus transfer from irish grassland
  soils to rivers—integration of laboratory and catchment studies.
\newblock {\em Journal of Hydrology}, 304(1):20--34.

\bibitem[Kang et~al., 2014]{Kang2014}
Kang, S.-M., Radhakrishnan, R., You, Y.-H., Joo, G.-J., Lee, I.-J., Lee, K.-E.,
  and Kim, J.-H. (2014).
\newblock Phosphate solubilizing bacillus megaterium mj1212 regulates
  endogenous plant carbohydrates and amino acids contents to promote mustard
  plant growth.
\newblock {\em Indian Journal of Microbiology}, 54(4):427--433.

\bibitem[Khan et~al., 2010]{Khan2010}
Khan, M.~S., Zaidi, A., Ahemad, M., Oves, M., and Wani, P.~A. (2010).
\newblock Plant growth promotion by phosphate solubilizing fungi--current
  perspective.
\newblock {\em Archives of Agronomy and Soil Science}, 56(1):73--98.

\bibitem[Khan et~al., 2009]{Khan2009}
Khan, M.~S., Zaidi, A., Wani, P.~A., and Oves, M. (2009).
\newblock Role of plant growth promoting rhizobacteria in the remediation of
  metal contaminated soils.
\newblock {\em Environmental Chemistry Letters}, 7(1):1--19.

\bibitem[Kouas et~al., 2005]{Kouas2005}
Kouas, S., Labidi, N., Debez, A., and Abdelly, C. (2005).
\newblock Effect of {P} on nodule formation and {N} fixation in bean.
\newblock {\em Agronomy for Sustainable Development}, 25(3):389--393.

\bibitem[Krishnaraj et~al., 2014]{Krishnaraj2014}
Krishnaraj, P., Dahale, S., et~al. (2014).
\newblock Mineral phosphate solubilization: concepts and prospects in
  sustainable agriculture.
\newblock In {\em Proc Indian Natl Sci Acad}, volume~80, pages 389--405.

\bibitem[Langergraber et~al., 2009]{Langergraber2009}
Langergraber, G., Rousseau, D. P.~L., Garc{\'{i}}a, J., and Mena, J. (2009).
\newblock {CWM1: A general model to describe biokinetic processes in subsurface
  flow constructed wetlands}.
\newblock {\em Water Science and Technology}, 59(9):1687--1697.

\bibitem[Louca and Doebeli, 2016]{Louca2016}
Louca, S. and Doebeli, M. (2016).
\newblock Reaction-centric modeling of microbial ecosystems.
\newblock {\em Ecological Modelling}, 335:74--86.

\bibitem[Macklon et~al., 1997]{Macklon1997}
Macklon, A., Grayston, S., Shand, C., Sim, A., Sellars, S., and Ord, B. (1997).
\newblock Uptake and transport of phosphorus by agrostis capillaris seedlings
  from rapidly hydrolysed organic sources extracted from 32 p-labelled
  bacterial cultures.
\newblock {\em Plant and Soil}, 190(1):163--167.

\bibitem[Maggi et~al., 2008]{Maggi2008}
Maggi, F., Gu, C., Riley, W., Hornberger, G., Venterea, R., Xu, T., Spycher,
  N., Steefel, C., Miller, N., and Oldenburg, C. (2008).
\newblock A mechanistic treatment of the dominant soil nitrogen cycling
  processes: Model development, testing, and application.
\newblock {\em Journal of Geophysical Research: Biogeosciences}, 113(G2).

\bibitem[Maier et~al., 2009]{Maier2009}
Maier, R.~M., Pepper, I.~L., and Gerba, C.~P. (2009).
\newblock {\em Environmental microbiology}, volume 397.
\newblock Academic press.

\bibitem[McGroddy et~al., 2004]{Mcgroddy2004}
McGroddy, M.~E., Daufresne, T., and Hedin, L.~O. (2004).
\newblock Scaling of {C}: {N}: {P} stoichiometry in forests worldwide:
  Implications of terrestrial redfield-type ratios.
\newblock {\em Ecology}, 85(9):2390--2401.

\bibitem[McGuinness et~al., 2014]{Mcguinness2014}
McGuinness, M., Cribbin, L., Winstanley, H., and Fowler, A. (2014).
\newblock Modelling spatial oscillations in soil borehole bacteria.
\newblock {\em Journal of Theoretical Biology}, 363:74--79.

\bibitem[Meena et~al., 2016]{Meena2016}
Meena, V.~S., Maurya, B., Meena, S.~K., Meena, R.~K., Kumar, A., Verma, J., and
  Singh, N. (2016).
\newblock Can bacillus species enhance nutrient availability in agricultural
  soils?
\newblock In {\em Bacilli and Agrobiotechnology}, pages 367--395. Springer.

\bibitem[Monod, 1949]{Monod1949}
Monod, J. (1949).
\newblock The growth of bacterial cultures.
\newblock {\em Annual Reviews in Microbiology}, 3(1):371--394.

\bibitem[Moyles and Fowler, 2018]{Moyles2018}
Moyles, I.~R. and Fowler, A.~C. (2018).
\newblock Production of nitrate spikes in a model of ammonium biodegradation.
\newblock {\em Theoretical Ecology}, 11(3):333--350.

\bibitem[Oehl et~al., 2001]{Oehl2001}
Oehl, F., Oberson, A., Probst, M., Fliessbach, A., Roth, H.-R., and Frossard,
  E. (2001).
\newblock Kinetics of microbial phosphorus uptake in cultivated soils.
\newblock {\em Biology and Fertility of Soils}, 34(1):31--41.

\bibitem[Pan et~al., 2014]{Pan2014}
Pan, Y., Cassman, N., de~Hollander, M., Mendes, L.~W., Korevaar, H., Geerts,
  R.~H., van Veen, J.~A., and Kuramae, E.~E. (2014).
\newblock Impact of long-term {N}, {P}, {K}, and {NPK} fertilization on the
  composition and potential functions of the bacterial community in grassland
  soil.
\newblock {\em FEMS Microbiology Ecology}, 90(1):195--205.

\bibitem[Ramirez et~al., 2012]{Ramirez2012}
Ramirez, K.~S., Craine, J.~M., and Fierer, N. (2012).
\newblock Consistent effects of nitrogen amendments on soil microbial
  communities and processes across biomes.
\newblock {\em Global Change Biology}, 18(6):1918--1927.

\bibitem[Redfield, 1934]{Redfield1934}
Redfield, A.~C. (1934).
\newblock On the proportions of organic derivatives in sea water and their
  relation to the composition of plankton.
\newblock {\em James Johnstone Memorial Volume}, pages 176--192.

\bibitem[Reich and Oleksyn, 2004]{Reich2004}
Reich, P.~B. and Oleksyn, J. (2004).
\newblock Global patterns of plant leaf {N} and {P} in relation to temperature
  and latitude.
\newblock {\em Proceedings of the National Academy of Sciences},
  101(30):11001--11006.

\bibitem[Richardson et~al., 2009]{Richardson2009}
Richardson, A.~E., Barea, J.-M., McNeill, A.~M., and Prigent-Combaret, C.
  (2009).
\newblock Acquisition of phosphorus and nitrogen in the rhizosphere and plant
  growth promotion by microorganisms.
\newblock {\em Plant and Soil}, 321(1-2):305--339.

\bibitem[Richardson and Simpson, 2011]{Richardson2011}
Richardson, A.~E. and Simpson, R.~J. (2011).
\newblock Soil microorganisms mediating phosphorus availability update on
  microbial phosphorus.
\newblock {\em Plant Physiology}, 156(3):989--996.

\bibitem[Rodr{\i}guez and Fraga, 1999]{Rodriguez1999}
Rodr{\i}guez, H. and Fraga, R. (1999).
\newblock Phosphate solubilizing bacteria and their role in plant growth
  promotion.
\newblock {\em Biotechnology Advances}, 17(4):319--339.

\bibitem[Roose et~al., 2001]{Roose2001}
Roose, T., Fowler, A., and Darrah, P. (2001).
\newblock A mathematical model of plant nutrient uptake.
\newblock {\em Journal of mathematical biology}, 42(4):347--360.

\bibitem[Rosenberg et~al., 1977]{Rosenberg1977}
Rosenberg, H., Gerdes, R., and Chegwidden, K. (1977).
\newblock Two systems for the uptake of phosphate in escherichia coli.
\newblock {\em Journal of Bacteriology}, 131(2):505--511.

\bibitem[Runyan and D'Odorico, 2012]{Runyan2012}
Runyan, C. and D'Odorico, P. (2012).
\newblock Hydrologic controls on phosphorus dynamics: A modeling framework.
\newblock {\em Advances in Water Resources}, 35:94--109.

\bibitem[Runyan and D'Odorico, 2013]{Runyan2013}
Runyan, C. and D'Odorico, P. (2013).
\newblock Positive feedbacks and bistability associated with
  phosphorus--vegetation--microbial interactions.
\newblock {\em Advances in Water Resources}, 52:151--164.

\bibitem[Saeid et~al., 2018]{Saeid2018}
Saeid, A., Prochownik, E., and Dobrowolska-Iwanek, J. (2018).
\newblock Phosphorus solubilization by bacillus species.
\newblock {\em Molecules}, 23(11):2897.

\bibitem[Saggar et~al., 1996]{Saggar1996}
Saggar, S., Parshotam, A., Sparling, G., Feltham, C., and Hart, P. (1996).
\newblock 14{C}-labelled ryegrass turnover and residence times in soils varying
  in clay content and mineralogy.
\newblock {\em Soil Biology and Biochemistry}, 28(12):1677--1686.

\bibitem[Santos et~al., 2011]{Santos2011}
Santos, H.~C., Oliveira, F. H. T.~d., Salcedo, I.~H., Souza, A. P.~d., Silva,
  M., et~al. (2011).
\newblock Kinetics of phosphorus sorption in soils in the state of
  para{\'\i}ba.
\newblock {\em Revista Brasileira de Ci{\^e}ncia do Solo}, 35(4):1301--1310.

\bibitem[Schachtman et~al., 1998]{Schachtman1998}
Schachtman, D.~P., Reid, R.~J., and Ayling, S.~M. (1998).
\newblock Phosphorus uptake by plants: from soil to cell.
\newblock {\em Plant Physiology}, 116(2):447--453.

\bibitem[Seeling and Zasoski, 1993]{Seeling1993}
Seeling, B. and Zasoski, R.~J. (1993).
\newblock Microbial effects in maintaining organic and inorganic solution
  phosphorus concentrations in a grassland topsoil.
\newblock {\em Plant and Soil}, 148(2):277--284.

\bibitem[Sharma et~al., 2013]{Sharma2013}
Sharma, S.~B., Sayyed, R.~Z., Trivedi, M.~H., and Gobi, T.~A. (2013).
\newblock Phosphate solubilizing microbes: sustainable approach for managing
  phosphorus deficiency in agricultural soils.
\newblock {\em SpringerPlus}, 2(1):587.

\bibitem[Slomp and Cappellen, 2007]{Slomp2007}
Slomp, C. and Cappellen, P.~V. (2007).
\newblock The global marine phosphorus cycle: sensitivity to oceanic
  circulation.
\newblock {\em Biogeosciences}, 4(2):155--171.

\bibitem[Swain et~al., 2012]{Swain2012}
Swain, M.~R., Laxminarayana, K., and Ray, R.~C. (2012).
\newblock Phosphorus solubilization by thermotolerant bacillus subtilis
  isolated from cow dung microflora.
\newblock {\em Agricultural Research}, 1(3):273--279.

\bibitem[Van~Kauwenbergh, 2010]{VanKauwenbergh2010}
Van~Kauwenbergh, S.~J. (2010).
\newblock {\em World phosphate rock reserves and resources}.
\newblock IFDC Muscle Shoals.

\bibitem[Vershinina and Znamenskaya, 2002]{Vershinina2002}
Vershinina, O. and Znamenskaya, L. (2002).
\newblock The pho regulons of bacteria.
\newblock {\em Microbiology}, 71(5):497--511.

\bibitem[Voegele et~al., 1997]{Voegele1997}
Voegele, R.~T., Bardin, S., and Finan, T.~M. (1997).
\newblock Characterization of the rhizobium (sinorhizobium) meliloti high-and
  low-affinity phosphate uptake systems.
\newblock {\em Journal of Bacteriology}, 179(23):7226--7232.

\bibitem[Vrede et~al., 2002]{Vrede2002}
Vrede, K., Heldal, M., Norland, S., and Bratbak, G. (2002).
\newblock Elemental composition ({C}, {N}, {P}) and cell volume of
  exponentially growing and nutrient-limited bacterioplankton.
\newblock {\em Applied Environmental Microbiology}, 68(6):2965--2971.

\bibitem[Zou et~al., 1992]{Zou1992}
Zou, X., Binkley, D., and Doxtader, K.~G. (1992).
\newblock A new method for estimating gross phosphorus mineralization and
  immobilization rates in soils.
\newblock {\em Plant and Soil}, 147(2):243--250.

\end{thebibliography}

\end{document}